2023-11-13

# Sub-Doppler optical-optical double-resonance spectroscopy using a cavity-enhanced frequency comb probe

**Vinicius Silva de Oliveira[1], Isak Silander[1], Lucile Rutkowski[2], Grzegorz Soboń[3], Ove Axner[1], Kevin K. Lehmann[4], and Aleksandra Foltynowicz[1,*]**

[1] Department of Physics, Umeå University, 901 87 Umeå, Sweden

[2] Univ Rennes, CNRS, IPR (Institut de Physique de Rennes)-UMR 6251, F-35000 Rennes, France

[3] Faculty of Electronics, Photonics and Microsystems, Wrocław University of Science and Technology, Wybrzeże Wyspiańskiego 27, 50-370 Wrocław, Poland

[4] Departments of Chemistry & Physics, University of Virginia, Charlottesville, VA 22904, USA

Corresponding author: aleksandra.foltynowicz@umu.se

**Abstract:** Accurate parameters of molecular hot-band transitions, *i.e.*, those starting from vibrationally excited levels, are needed to accurately model high-temperature spectra in astrophysics and combustion, yet laboratory spectra measured at high temperatures are often unresolved and difficult to assign. Optical-optical double-resonance (OODR) spectroscopy allows the measurement and assignment of individual hot-band transitions from selectively pumped energy levels without the need to heat the sample. However, previous demonstrations lacked either sufficient resolution, spectral coverage, absorption sensitivity, or frequency accuracy. Here we demonstrate OODR spectroscopy using a cavity-enhanced frequency comb probe that combines all these advantages. We detect and assign sub-Doppler transitions in the spectral range of the $3\nu_3 \leftarrow \nu_3$ resonance of methane with frequency precision and sensitivity more than an order of magnitude better than before. This technique will provide high-accuracy data about excited states of a wide range of molecules that is urgently needed for theoretical modeling of high-temperature data and cannot be obtained using other methods.

## Introduction

Absorption spectroscopy is one of a few techniques that allow *in-situ* analysis of high-temperature molecular gases, with applications including combustion science[1-3] and atmospheric sensing of hot astrophysical objects[4,5]. Identification of the species and their densities from absorption measurements requires accurate theoretical models of the molecular ro-vibrational energy structure that are verified using high-precision laboratory spectra. The launch of the James Webb Space Telescope (JWST) has placed in the spotlight the immediate need of rotationally resolved reference spectroscopic data for many small organic species in a wide range of thermodynamic conditions[5]. The first JWST survey of the VHS-1256b exoplanet[6] yielded observational evidence of atmospheric $H_2O$, CO, $CO_2$, and $CH_4$. These unambiguous detections were based on a comparison of the detected spectral features to synthetic spectra computed at the relevant temperature (1000 K). However, these models do not account for all observed transitions and many features remained unidentified. Recently, $CH_3^+$, a radical cation related to methane photochemistry, has been identified in a hot protoplanetary disk[7] by comparing its emission spectrum recorded by the JWST to a model spectrum developed based on available spectroscopic constants. The remaining discrepancies between the model and the spectrum imply either the presence of other species or inaccuracies in the model[7]. Resolving this question requires verification of the theoretical model using experimental data, which is not available.

Such unresolved detections at high temperatures underline the urgent need for precision measurements of molecular excited states to provide the spectroscopic parameters required to model the hot bands. Among the different species, methane poses a particular challenge. The fundamental vibrational mode frequencies of methane are nearly resonant and coupled through a number of strong interactions. The high density of excited levels and the strong couplings between them make them difficult to calculate using a*b-initio* methods[8]. State-of-the-art synthetic high-temperature line lists of methane are contained in, e.g., the TheoReTS[9] and the ExoMol databases[10]. The TheoReTS data have recently been incorporated in the HITEMP database[11], which is used as a reference in many high-temperature applications. However, the energy levels above 8000 $cm^{-1}$ – relevant to hot environments (>500 K) – remain largely unverified. Room- and low-temperature precision spectroscopy of overtone bands provides valuable information about levels that can be reached from the ground vibrational level[12] but often does not shed light on levels involved in hot-band transitions (*i.e.*, transitions starting from excited vibrational states). Obtaining empirical hot-band line lists from laboratory measurements is difficult because absorption and emission spectra measured at high temperatures are often congested with overlapping transitions, making them difficult to resolve and assign[13-15].

Optical-optical double-resonance (OODR) spectroscopy is a method that allows selective measurement of hot-band transitions without the need to heat the sample. In OODR, a strong pump laser populates a selected excited state, and a weaker probe laser measures hot-band transitions from this state, which results in a much less congested spectrum, simpler to analyze than a spectrum from a thermally excited sample. OODR spectroscopy has historically been performed using tunable pulsed lasers[16] that provide broad spectral coverage, but have limited spectral resolution and frequency accuracy, which often prevents resolving individual transitions. Compared to that, OODR spectroscopy using narrow-linewidth continuous-wave (CW) lasers has a number of advantages[17,18]: sub-Doppler resolution, because a pump with narrow linewidth excites only one velocity group of molecules; high absorption sensitivity, especially when combined with cavity-enhanced methods; and kHz frequency accuracy when the pump and probe lasers are referenced to a frequency comb. However, the tunability of narrow-linewidth CW lasers is limited and surveying large spectral ranges is time consuming, often making the search for transitions impractical and cumbersome.

Using frequency combs as probes in OODR spectroscopy overcomes these limitations and provides spectra with broad bandwidth, inherent absolute frequency calibration, and sub-Doppler resolution, allowing unambiguous detection of many hot-band transitions simultaneously. OODR based on a CW pump and a frequency comb probe was first performed on an atomic Rb sample[19,20] contained in a single-pass cell. This was possible because atomic transitions are 3 to 4 orders of magnitude stronger than molecular ones. Recently, we demonstrated OODR spectroscopy on a molecular sample[21] and used it to measure and assign 36 sub-Doppler transitions in the $3\nu_3 \leftarrow \nu_3$ resonance region of methane in probe spectra spanning 6 THz of bandwidth[22]. These results provided the first high-accuracy verification of theoretical predictions of hot-band transitions of methane in this range, finding better agreement with TheoReTS than ExoMol. However, the absorption sensitivity and frequency accuracy of these measurements were limited by the use of a single-pass cell for the methane sample. Even though the cell was cooled by liquid nitrogen to increase the intensity of the OODR probe transitions (by increasing the molecular population in the pumped low rotational states), the signal-to-noise ratio (SNR) of the OODR signal was at most 10. The requirement of cooling limited the applicability of the technique to methane, which is the only stable polyatomic molecule that has sufficient vapor pressure at 77 K. Moreover, in the cell, the pump and probe beams were co-propagating, and thus interacting with the same velocity group of molecules, so a residual drift of the pump laser frequency translated to a proportional shift of the probe transition frequencies, which limited the frequency accuracy.

Here, we introduce OODR spectroscopy using a cavity-enhanced frequency comb as the probe, which dramatically increases the absorption sensitivity and frequency precision of detection of hot-band transitions without the requirement of cooling of the sample, making it applicable to a large range of molecules. The cavity increases the interaction length of the probe with the sample, which allows detection of transitions that are more than an order of magnitude weaker than previously observed. Moreover, while the pump beam makes a single pass through the sample, the cavity-enhanced probe beam is both co- and counter-propagating with respect to the pump and simultaneously interacts with two molecular velocity groups with opposite signs. This cancels the influence of the pump frequency drift on the position of the probe lines, and improves the frequency precision by more than an order of magnitude. The high SNR and frequency precision allow using two independent methods of assigning the rotational quantum number of the final state of the probe transitions without the need to rely on theoretical predictions. The first is based on the differences of OODR probe line intensities measured with parallel and perpendicular relative pump/probe polarizations, which arise from the sample birefringence induced by the pump laser. The second is using combination differences, *i.e.*, reaching the same final state by different combinations of pump and probe frequencies. We show that these two methods are in agreement with each other, and the assignments are confirmed by theoretical predictions from the TheoReTS/HITEMP database. This opens up the sensitive detection and unambiguous assignment of hot-band transitions of molecules for which theoretical predictions are missing or inaccurate.

## Results

### *Experimental setup and procedures*

To demonstrate cavity-enhanced frequency comb OODR spectroscopy we use a high-power 3.3 μm (3000 $cm^{-1}$) CW pump and a 1.68 μm (5950 $cm^{-1}$)-centered frequency comb probe, as shown in Fig. 1(a). The pump frequency is Lamb-dip locked to a $CH_4$ transition from a vibrational ground state and populates a selected assigned state in the $\nu_3$ band, while the comb simultaneously probes sub-Doppler hot-band transitions from the pumped $\nu_3$ state and Doppler-broadened transitions from the ground state, as shown in Fig. 1(b). The final states reached by the sub-Doppler probe transitions have term values in the 8990 – 9010 $cm^{-1}$ range and belong to different sub-bands within the triacontad polyad of methane, where the dominating band is $3\nu_3 \leftarrow \nu_3$. The Doppler-broadened absorption is dominated by the $2\nu_3$ cold band.

The sample of 50 mTorr of pure $CH_4$ is contained in an 80-cm-long cavity resonant with the comb probe (187 MHz free spectral range, *FSR*) but highly transmitting and thus non-resonant with the pump. The cavity finesse varies between 5000 and 8500 in the 1672-1692 nm (5910-5980 $cm^{-1}$) range (see Supplementary information 1). The comb is locked to the cavity using the two-point Pound-Drever-Hall (PDH) stabilization scheme[23] (see Methods for details). The repetition rate, $f_{rep}$, is locked to a tunable source referenced to a GPS-disciplined Rb oscillator, and the carrier-envelope offset frequency, $f_{ceo}$, is monitored using an $f$-$2f$ interferometer. The pump and probe beams are combined in front of the cavity using a dichroic mirror and their relative polarization is adjusted to be either parallel or perpendicular using a half-wave plate in the pump beam. The pump passes once through the cavity and the transmitted power is monitored using a power meter after a second dichroic mirror that separates the pump and probe beams after the cavity.

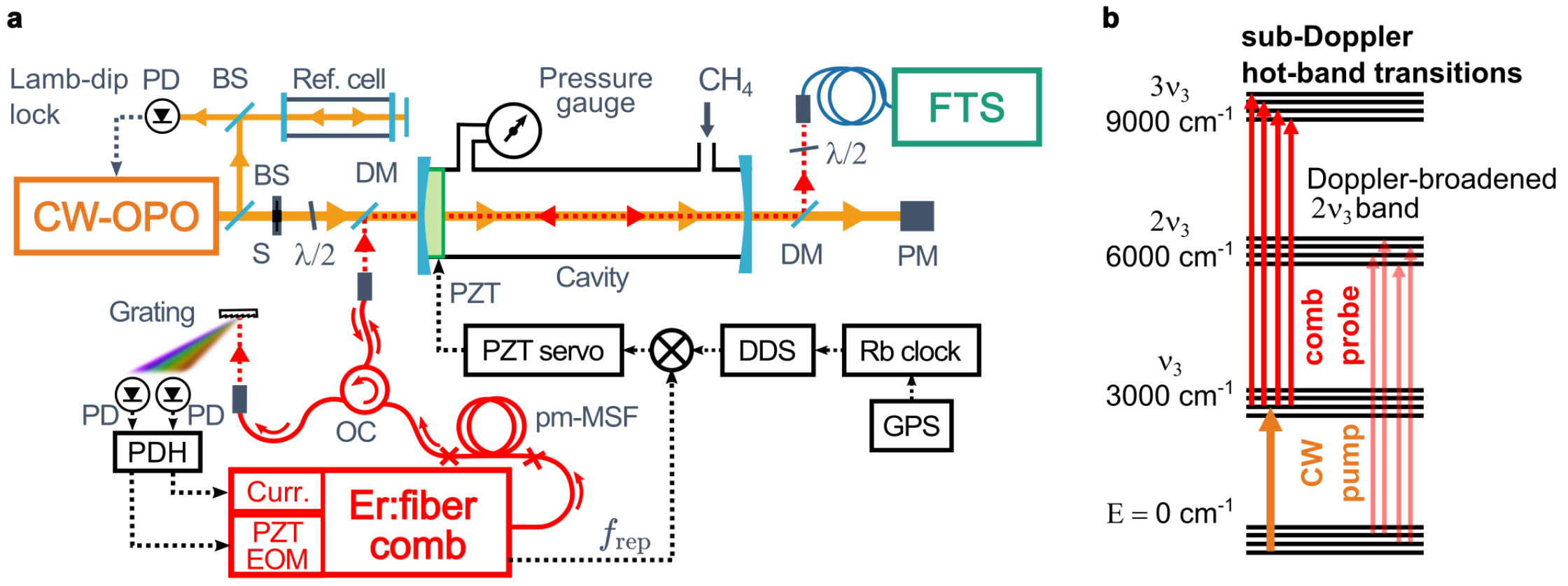

Fig. 1 **Energy levels involved in OODR spectroscopy and the experimental setup. a** Experimental setup. CW-OPO: continuous-wave optical parametric oscillator. BS: beam splitter. S: shutter. λ/2: half-wave plate. DM: dichroic mirrors. PZT: piezoelectric transducer. FTS: Fourier transform spectrometer. PM: power meter. pm-MSF: polarization-maintaining microstructured silica fiber. OC: optical circulator. PD: photodiodes. Curr.: current input of the comb oscillator. EOM: electro-optic modulator. DDS: direct digital synthesizer. **b** Simplified representation of the vibrational bands of methane addressed by the CW pump (orange) and the comb probe (red).

The transmitted comb has a bandwidth of 1.5 THz (15 nm, 50 cm$^{-1}$) limited by cavity mirror dispersion, and its center frequency can be tuned anywhere within the bandwidth of the incident comb by a proper choice of the locking points in the two-point PDH scheme. The comb beam is led via a polarization-maintaining optical fiber to a fast-scanning Fourier transform spectrometer with auto-balanced detection[24]. Spectra are recorded at different $f_{rep}$ values and interleaved to yield a sample point spacing of 2 MHz in the optical domain (see Methods for details). Comb-mode-limited resolution is obtained using the method of Refs. [25,26], which relies on matching the nominal resolution of the spectrometer to the $f_{rep}$ (see Supplementary information 4). To remove the background originating from the comb envelope and the Doppler-broadened absorption of the 2ν$_3$ band, at each $f_{rep}$ step we use a shutter to acquire spectra with and without the CW pump excitation and take their ratio. The slowly varying baseline in the normalized spectrum, arising from intensity fluctuations of the cavity transmission, is removed using the cepstral method[27]. The total acquisition time of one normalized and interleaved spectrum containing 750,000 sampling points spaced by 2 MHz is 16.7 min.

We recorded spectra with the pump consecutively locked to three transitions in the ν$_3$ band starting from the same level in the ground state with rotational quantum number $J = 2$, namely the P(2, $F_2$), Q(2, $F_2$), and R(2, $F_2$) transitions. For the P(2, $F_2$) and Q(2, $F_2$) pump transitions, we recorded 5 series of $f_{rep}$ scans with both pump polarizations, while for the R(2, $F_2$) pump transition we recorded 5 series with perpendicular polarization, and 45 series with parallel polarization (see Supplementary information 2). The pump frequencies and the corresponding comb probe coverage are summarized in Table I. The OODR probe transitions were found in each interleaved spectrum using a peak detection routine similar to that used in Ref. [22].

Table I. **Pump transitions and probed ranges.** The transitions pumped in the three measurement series, their wavenumbers from Refs. [31,35] and the corresponding spectral coverage of the comb probe.

| Measurement | Pump transition ($\nu_3$ band) | Pump wavenumber[31,35] [$cm^{-1}$] | Probe coverage [$cm^{-1}$] |
|---|---|---|---|
| 1 | P(2, $F_2$) | 2998.99403200(7) | 5935 - 5985 |
| 2 | Q(2, $F_2$) | 3018.65020715(7) | 5925 - 5975 |
| 3 | R(2, $F_2$) | 3048.15331810(8) | 5905 - 5945 |

### *Sensitivity*

A narrow section of the interleaved and normalized probe spectrum recorded with the pump locked to the $\nu_3$ R(2, $F_2$) transition and averaged 5 times is shown by the black curve in Fig. 2, revealing two sub-Doppler OODR probe transitions on a flat baseline. The noise on the baseline is on average $\sigma = 4.7 \times 10^{-3}$, which translates to a lowest detectable absorption coefficient, $\alpha_{min} = \sigma/L_{eff}$, of $1.5 \times 10^{-8}$ $cm^{-1}$ at 1.4 h, where the effective cavity length, $L_{eff}$, is given by $2FL/\pi$, where, in turn, $L$ is the cavity length, and $F$ is the cavity finesse equal to 6000 at 5929 $cm^{-1}$, while 1.4 h is the measurement time of 5 interleaved spectra, $\tau$. The noise-equivalent absorption sensitivity, defined as $\alpha_{min}\tau^{1/2}$, is $1.1 \times 10^{-6}$ $cm^{-1}$ $Hz^{-1/2}$, which is a factor of 700 better than in the previous measurement employing the liquid-nitrogen-cooled single-pass cell[21,22]. At 110 K, the temperature of the previous single-pass-cell measurement, the pump signal for $J = 2$ lines at a given pressure is 14.6 times stronger than at room temperature (see Supplementary information 5). This implies that the room-temperature cavity allows detection of 700/14.6 = 50 times weaker probe transitions than before at the same pressure and acquisition time.

The figure of merit, defined as $\alpha_{min}(\tau/M)^{1/2}$, where $M$ is the number of spectral elements, is $1.3 \times 10^{-9}$ $cm^{-1}$ $Hz^{-1/2}$ per spectral element, on par with what previously has been achieved in the same spectral region using cavity-enhanced dual-comb spectroscopy[28] or in the mid-infrared using cavity-enhanced comb-based dispersive spectrometer designed for physical-chemistry applications[29]. However, none of the previously demonstrated cavity-enhanced comb-based spectrometers had sub-Doppler resolution and the capability to detect hot-band transitions.

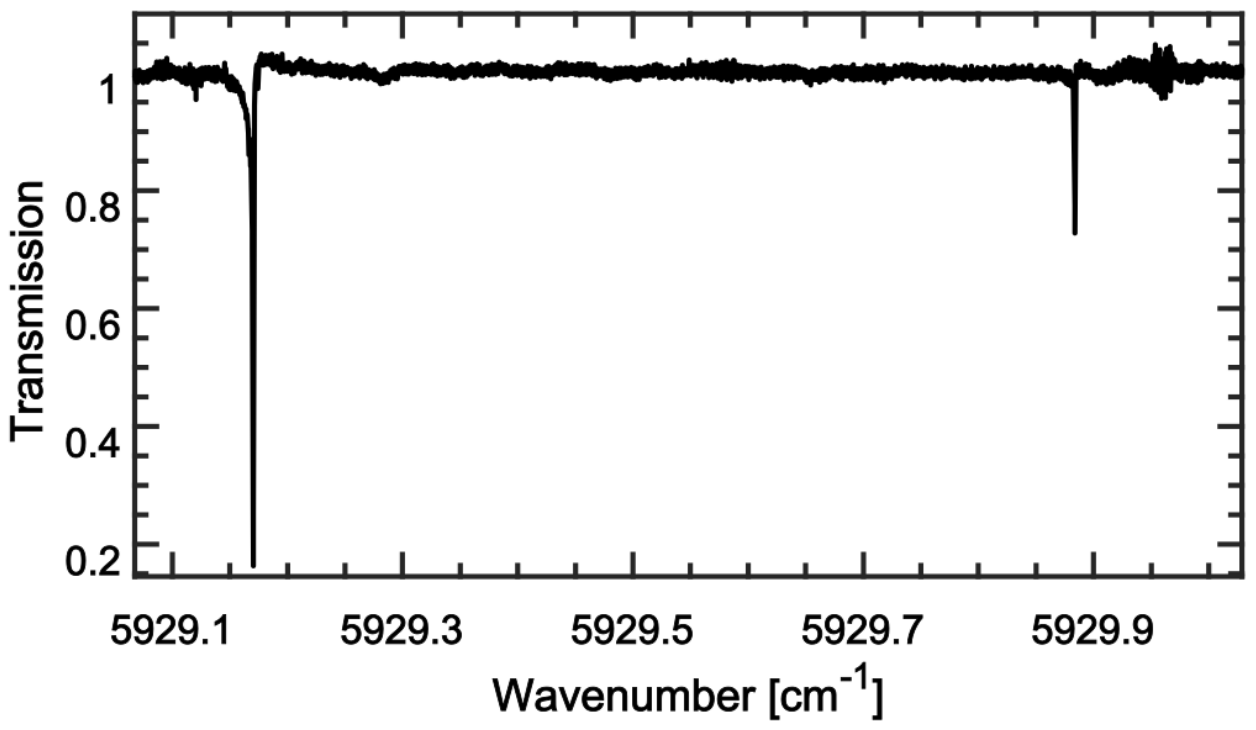

Fig. 2. **OODR spectrum.** A narrow section of the probe spectrum measured with the pump locked to the $\nu_3$ R(2, $F_2$) transition and perpendicular relative pump/probe polarizations (5 averages) that contains two sub-Doppler probe transitions.

### *Frequency uncertainty*

The three strongest OODR probe transitions detected with the pump locked to the $\nu_3$ R(2, $F_2$) transition are shown in Fig. 3 for parallel (red markers) and perpendicular (black markers) relative pump/probe polarizations. The curves show fits of the cavity-enhanced transmission function[23] (see Supplementary information 3), from which we retrieve the center frequencies, integrated absorptions, and widths of the probe lines. The asymmetry in the line shapes, visible in Fig. 3(a) and (b), is caused by the offset of the comb modes from cavity resonances, which in turn is caused primarily by dispersion of the cavity mirror coatings. This comb-cavity offset is zero close to the PDH locking points, e.g. in Fig. 3(c), and increases away from them. This effect is included in the cavity transmission function and does not affect the accuracy of the center frequency determination. The residuals visible around the line centers indicate that modeling the OODR probe transitions as single Lorentzian peaks (see Supplementary information 3) is not fully appropriate, and work is ongoing on improving the accuracy of the model. We note that the residuals are symmetric around the line center and thus the inaccuracy of the model does not affect the accuracy of the center frequency determination. The width of the lines is of the order of 5 MHz, dominated by power broadening caused by the pump.

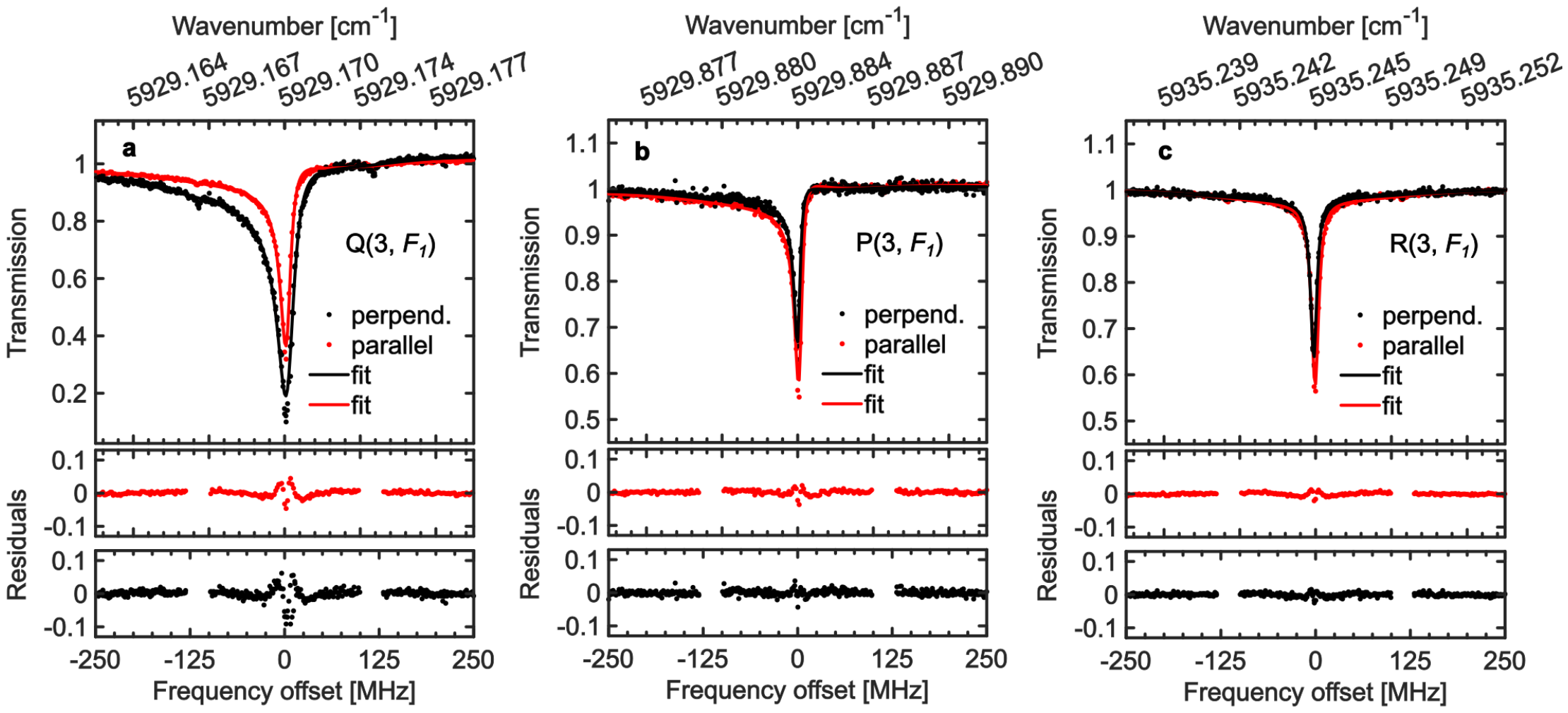

Fig. 3. **OODR probe transitions.** Three probe transitions: **a** $3\nu_3 \leftarrow \nu_3$ Q(3, $F_1$), **b** $\nu_1+4\nu_2 \leftarrow \nu_3$ P(3, $F_1$), and **c** $\nu_2+\nu_3+\nu_4 \leftarrow \nu_3$ R(3, $F_1$), measured when the pump is locked to the $\nu_3$ R(2, $F_2$) transition. Upper windows: data taken with parallel (red markers, 5 averages) and perpendicular (black markers, 5 averages) relative pump/probe polarizations together with fits of the cavity transmission function (solid curves). Lower windows: residuals of the fits. Line assignment – see text under Line assignments.

To investigate the long-term stability of center frequency, we performed fits to 9 OODR probe transitions detected in the 45 consecutive spectra recorded over 12 h with the pump locked to the $\nu_3$ R(2, $F_2$) transition and parallel relative pump/probe polarizations. The center frequencies of the three transitions from Fig. 3, obtained from these fits, are shown in Fig. 4, offset by their mean value. The error bars show the statistical standard errors from the individual fits, which vary between 15 and 112 kHz, while the shaded area indicates one standard deviation of all values, equal to 150 kHz ($5 \times 10^{-6}$ cm$^{-1}$). We attribute the fact that the spread of the center frequencies of the fits is larger than their precision to residual uncorrected baseline drift. The 150 kHz precision of the center frequency, for lines with SNR larger than 50 (see Supplementary information 4.2), is more than an order of magnitude better than obtained previously in the single-pass cell (1.7 MHz)[22], which confirms that the influence of the drift of the pump center frequency on the positions of the probe lines has been canceled.

We note that, based upon literature values for other $CH_4$ ro-vibrational transitions, the influence of the pressure and Stark shift on the probe transition frequencies is smaller than the precision of the measurement. Using the self-induced pressure shift coefficient of (–0.017 ± 0.003) cm$^{-1}$/atm$^{-1}$ for methane lines in the 6000 cm$^{-1}$ range reported by Lyulin et al.[30], yields a –33 kHz pressure shift for the probe lines at 50 mTorr, which is below the uncertainty we report. Okubo et al.[31] reported a power shift coefficient of a sub-Doppler Lamb dip in the P(7,$E$) line of the $\nu_3$ band to be (–13 ± 17) kHz/W for a beam radius of 0.71 mm. For the beam radius and power of pump in our experiment, this results in a (–0.8 ± 1.1) kHz shift, which also is negligible.

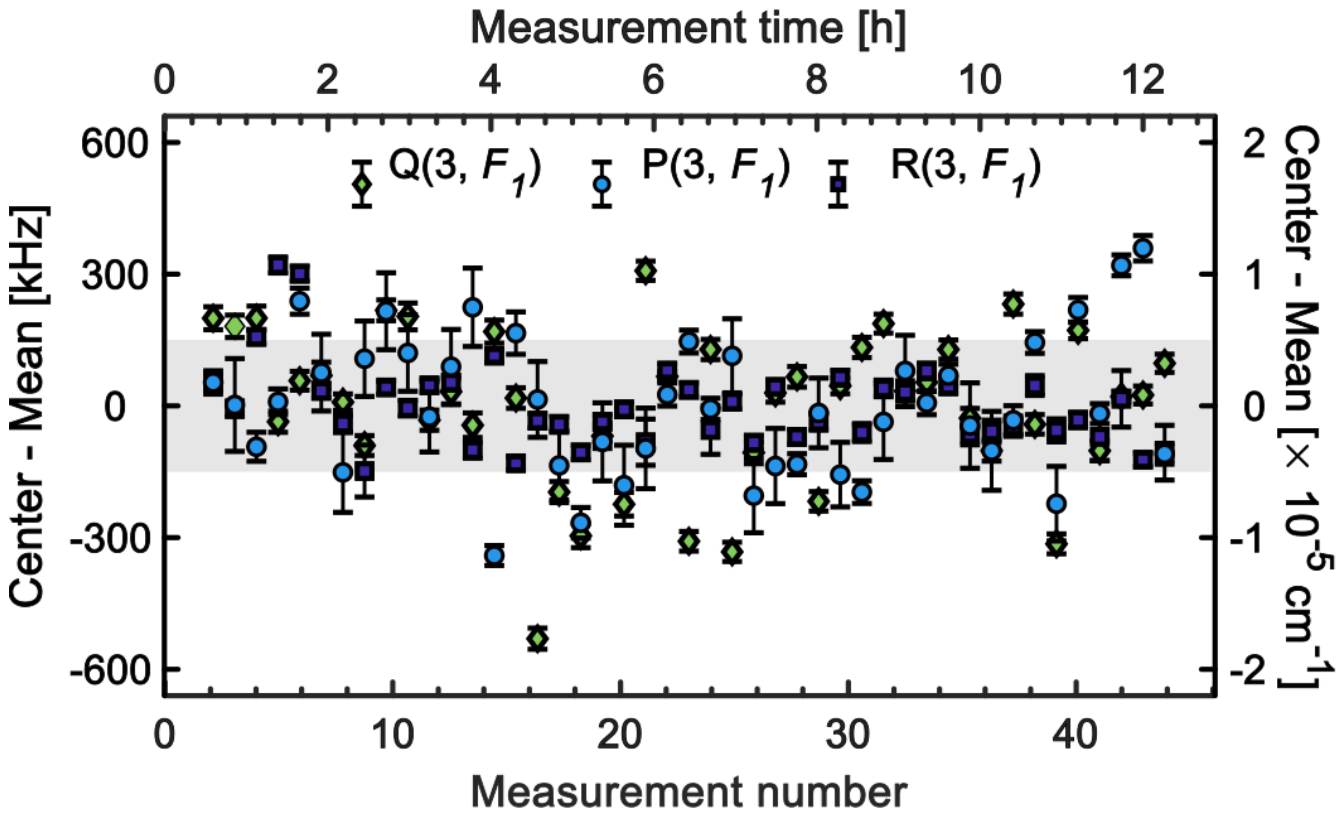

Fig. 4. **Long-term frequency measurement.** Center frequencies (left axis) and wavenumbers (right axis) from fits to 45 consecutive measurements of the three probe transitions shown in Fig. 3, as marked in the legend, offset by their mean. The error bars show the fit precisions, and the shaded area indicates one standard deviation of all values.

### *Line assignments*

We assign the branches of the detected OODR probe transitions using two independent methods. The first method uses the fact that the ratio of the probe line intensities measured with parallel and perpendicular relative pump/probe polarizations depends on the change of rotational quantum number $J$, *i.e.*, it is different for P, Q and R pump and probe transitions[32]. This is because the dipole moments of both the pump and the probe transitions, for each value of the projection of the total angular momentum on the quantization axis (defined by the pump electric field), depend on the total angular momentum quantum numbers of the two states of the transition and the direction of the optical electric field. The polarization-dependent intensity ratios can thus be predicted from the transition dipoles of the pump and probe transitions[32] (see Supplementary information 6). For example, when an R(2) transition is pumped, the parallel over perpendicular polarization integrated intensity ratios are predicted to be 1.85, 0.35, and 1.30 for a P(3), Q(3) and R(3) probe transition, respectively. For the lines shown in Fig. 3(a), 3(b) and 3(c), these intensity ratios are 0.4(1), 1.6(2), and 1.36(9), respectively, where the uncertainty is mainly given by the uncertainty in probe polarization (see Supplementary information 6). This suggests the line assignment as Q(3, $F_1$), P(3, $F_1$), and R(3, $F_1$). Table II lists in the last column the predicted and measured intensity ratios for all combinations of pump and probe transitions detected in this work.

The second method of branch assignment is based on combination differences, *i.e.*, cases when the same final energy state is reached by two or three different combinations of pump and probe frequencies, as is schematically shown in Fig. 5. Selection rules allow assigning the rotational quantum number $J$ of the final states depending on which probe spectra a given final state appears in. To find the combination differences among the measured transitions, we calculate their final state term value as the sum of the ground state term value 31.4423878(8) $cm^{-1}$ from Ref. [34], the pump transition frequencies from Refs. [31,35] (known with kHz accuracy, see Table I), and the measured probe transition wavenumbers (listed in Supplementary Table S2 in the Supplementary information 7). Common final states for different combinations of pump and probe transitions are easily identified as states whose term values agree within the experimental uncertainty, while the separations between the different final states are significantly larger that the experimental uncertainty. This confirms that the experimental uncertainties are not underestimated. Table II lists the final state term values reached by the three probe transitions shown in Fig. 3, together with other pump/probe combinations that reach the same states – if they exist. The probe transition assignment shown in column 3 is based on the combination differences, and it is consistent with the assignment based on the intensity ratios, shown in the last column of the table. The dominant assignment of the final state in column 6 is obtained from the non-empirical effective Hamiltonian described in Ref. [33].

We note that if transitions are not missed because of too low intensity, overlap with Doppler-broadened $2\nu_3$ transitions whose absorption is saturated, or because of being outside the probed spectral region, this method gives an unambiguous $J$ assignment for each final state reached.

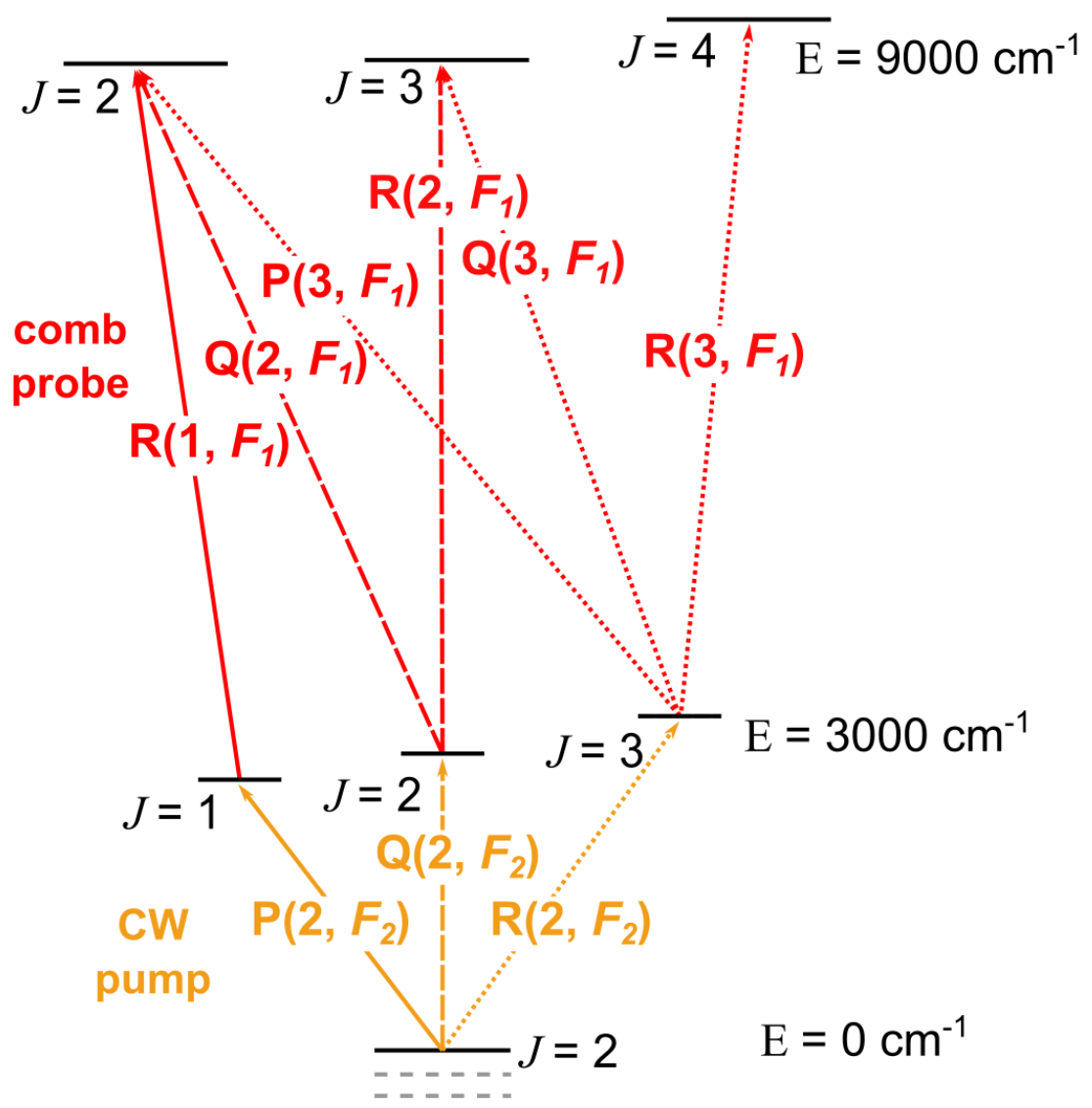

Fig. 5. **Final state assignment using combination differences.** Simplified illustration of the observed combination differences used to assign the final states of probe transitions. The solid, dashed, and dotted lines illustrate the transitions corresponding to the case when the P(2, $F_2$), Q(2, $F_2$), and R(2, $F_2$) transitions in the $\nu_3$ band are pumped, respectively. Transitions reaching final states with $J = 0$ and 1 are out of the measured probe range in this work and are not shown.

Table II: **Parameters of the OODR transitions.** Parameters of the transitions shown in Fig. 5 that allow assignment of the rotational quantum number of the final state via combination differences. Column 1: Pump transition. Column 2: Pump transition upper state term value, calculated as a sum of pump transition frequencies from Refs [31,35] and ground state term value from Ref. [34]. Column 3: Assignment of the probe transition based on combination differences. Column 4: Experimental probe transition wavenumber. Column 5: Final state term value. Column 6: Final state dominant assignment from the non-empirical effective Hamiltonian described in Ref. [33]. Column 7: Predicted (from Ref. [32]) and experimental intensity ratios for parallel and perpendicular relative pump/probe polarizations.

| Pump transition ($\nu_3$ band) | Pump transition upper state term value [cm$^{-1}$] | Probe transition | Probe transition wavenumber [cm$^{-1}$] | Final state term value [cm$^{-1}$] | Final state assignment[33] | Polarization-dependent intensity ratio | |
|---|---|---|---|---|---|---|---|
| | | | | | | predicted[32] | measured |
| P(2, $F_2$) | 3030.4364198(8) | R(1, $F_1$) | 5979.042972(4) | 9009.479391(4) | | 1.01 | 0.98(6) |
| Q(2, $F_2$) | 3050.0925950(8) | Q(2, $F_1$) | 5959.386794(5) | 9009.479389(5) | $\nu_1+4\nu_2$ ($A_1$) | 2.00 | 1.8(3) |
| R(2, $F_2$) | 3079.5957059(8) | P(3, $F_1$) | 5929.883689(4) | 9009.479394(4) | | 1.85 | 1.6(2) |
| | | | | | | | |
| Q(2, $F_2$) | 3050.0925950(8) | R(2, $F_1$) | 5958.673570(6) | 9008.766165(6) | $3\nu_3$ ($F_1$) | 0.80 | 0.83(7) |
| R(2, $F_2$) | 3079.5957059(8) | Q(3, $F_1$) | 5929.170466(4) | 9008.766172(4) | | 0.35 | 0.4(1) |
| | | | | | | | |
| R(2, $F_2$) | 3079.5957059(8) | R(3, $F_1$) | 5935.245195(4) | 9014.840901(4) | $3\nu_2+\nu_3+\nu_4$ ($F_2$) | 1.30 | 1.36(9) |

***Comparison to theoretical predictions***

In total, we detected 21 OODR probe transitions, whose intensities span more than two orders of magnitude, as shown in Fig. 6(a). 15 transitions, marked by rhombs, squares, and circles for the P(2, $F_2$), Q(2, $F_2$), and R(2, $F_2$)-pumped spectra, respectively, were measured with both relative pump/probe polarizations and could be assigned using the polarization-dependent intensity ratios, while the 6 weakest transitions, marked by stars, were observed only in the 45-times-averaged spectrum with pump locked to the $\nu_3$ R(2, $F_2$) transition and parallel pump/probe polarizations. For 9 transitions, the assignment is confirmed using combination differences.

We compare the measured line intensities and positions to predictions from the TheoReTS/HITEMP database[11]. Fig. 6(b) shows the ratios of the experimental and predicted integrated absorptions of the probe lines, while Fig. 6(c) displays the differences between the center wavenumbers of the observed and the predicted transitions (see Methods for details). We note that the TheoReTS predictions are missing for two weak probe transitions, marked by the gray stars in Fig. 6(a). The two outliers that are visible in the relative intensity plot, Fig. 6(b), correspond to two of the weakest detected lines, for which the predictions might be less accurate. Neglecting these two outliers, the mean intensity ratios for the lines in the P(2, $F_2$), Q(2, $F_2$) and R(2, $F_2$)-pumped spectra are constant to within 12%, 5% and 17%, respectively, while TheoReTS states an average accuracy of 2-3% on the integrated absorption[36]. The line positions are within 1.3 cm$^{-1}$ from the predictions, which is roughly within the estimated TheoReTS accuracy of 1 cm$^{-1}$.

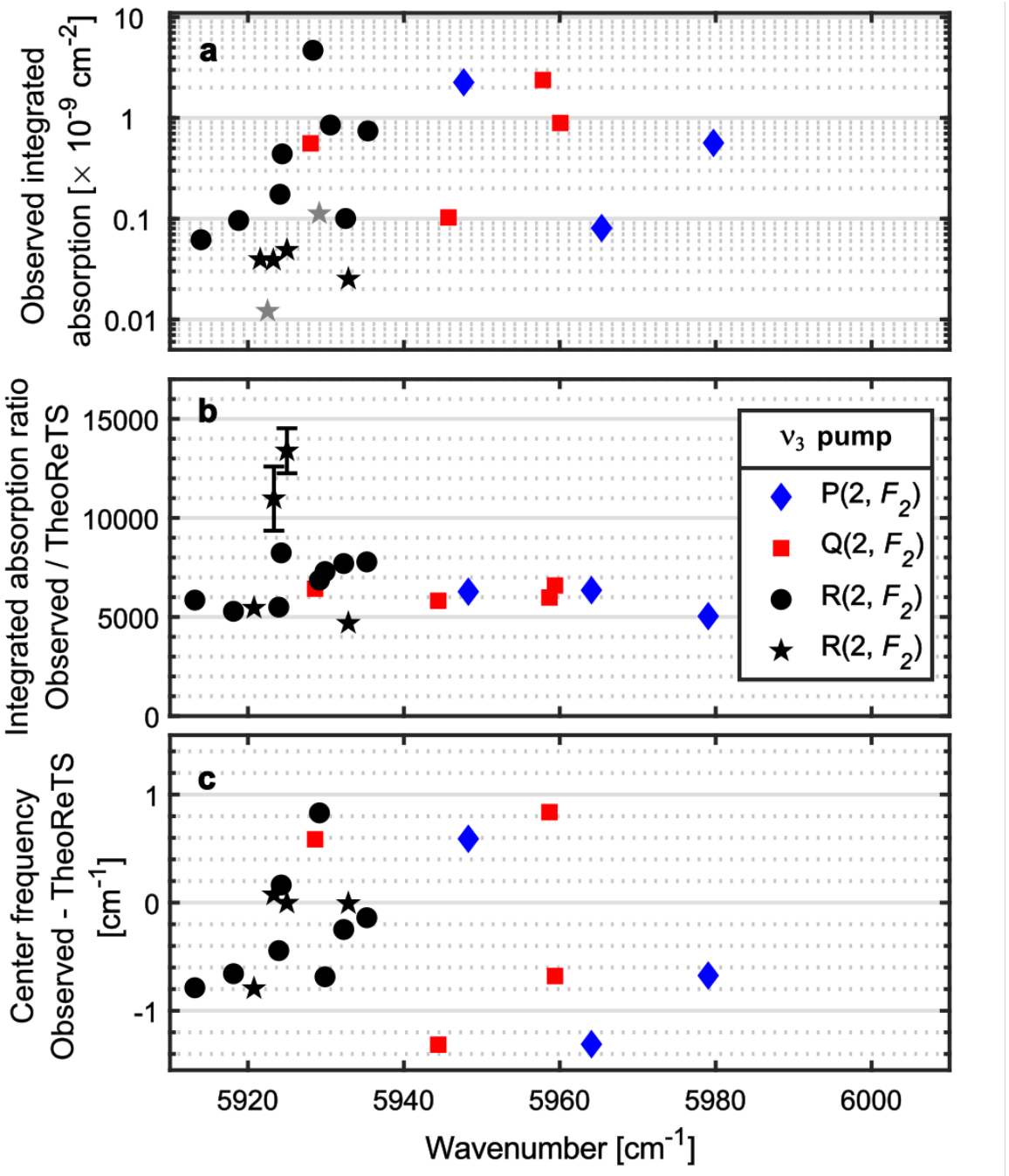

Fig. 6. **Comparison to the TheoReTS/HITEMP database.** **a** Integrated absorption of all measured OODR probe transitions on a logarithmic scale. **b** Ratios of the experimental integrated absorption from **a** and the integrated absorption predicted at 296 K and 50 mTorr using line intensities from the TheoReTS/HITEMP database. The error bars for all lines but two are of the order of the marker size and therefore not plotted. **c** Center wavenumbers of the OODR probe transitions compared to predictions from the TheoReTS/HITEMP database. The experimental uncertainties are negligible on this scale. The rhombs, squares, and circles correspond to the different pumped transitions in the $\nu_3$ band, as indicated in the legend. The stars indicate the 6 weak transitions found only in the dataset averaged 45 times with pump locked to the $\nu_3$ R(2, $F_2$) transition. The two transitions marked by gray stars in **a** lack assignment in TheoReTS/HITEMP and are therefore missing in **b** and **c**.

## Discussion and Outlook

In this work, we demonstrate cavity-enhanced frequency comb OODR spectroscopy that allows the detection and assignment of sub-Doppler hot-band transitions over a wide spectral range with high absorption sensitivity and frequency precision on the 150 kHz level. Compared to the previous demonstration of comb-based OODR that employed a liquid-nitrogen-cooled single-pass cell[21,22], the use of the cavity increases by more than an order of magnitude both the absorption sensitivity (by increasing the interaction length of the probe with the sample) and frequency precision (by eliminating the frequency shift caused by the residual drift in the pump frequency) while allowing operation at room temperature. The lack of cooling improves the accuracy of intensity measurements since there is a negligible temperature gradient in the cell. The room temperature operation will allow measurements of hot-band transitions from highly rotationally excited levels that are not accessible for the pump at liquid nitrogen temperatures, and for which theoretical predictions have not yet been verified. Most importantly, the technique can now be applied to molecules other than methane that would condense in a liquid-nitrogen-cooled cell.

The broad spectral coverage of the comb probe, the high frequency precision and the high SNR allow using two independent methods of assigning the rotational quantum number of the final states of the probe transitions, namely combination differences and polarization-dependent intensity ratios. The two methods are in agreement with each other, and the assignments are confirmed by predictions from the TheoReTS database. This implies that, in the future, it will be possible to assign transitions that lack theoretical predictions.

Cavity-enhanced comb-based OODR spectroscopy allows measuring and assigning individual hot-band transitions that would be indiscernible in high-temperature absorption spectra measured at local thermodynamic equilibrium. Using different combinations of CW pump and comb probe frequencies, the technique will enable systematic measurements and assignments of weak hot-band transitions of many molecules over a broad spectral range. The energy levels determined in this work are involved in hot-band transitions spanning the entire JWST observation window. Thus, an OODR measurement in one range has an impact on the accuracy of hot-band predictions across the entire infrared range covered by the JWST instruments. Reference data provided by this technique will lead to improved and new theoretical predictions of high-temperature spectra of many molecular species, needed to confirm (or disprove) their detections in future astrophysical observations.

## Methods

### *Pump and probe frequency stabilization and cavity mode matching*

The pump is the idler of a singly-resonant CW optical parametric oscillator (CW-OPO, Aculight, Argos 2400 SF, module C). Its frequency is stabilized to the center of the Lamb dip in the selected $CH_4$ transition in the $\nu_3$ band using a frequency-modulated error signal (modulation frequency 60 MHz) from a reference cell, as described in detail in Ref. [22]. The pressure in the Lamb dip cell was 30 mTorr for locking to the Q(2, $F_2$) and R(2, $F_2$) transitions, and 190 mTorr for locking to the weaker P(2, $F_2$) transition.

The probe is an amplified Er:fiber frequency comb (Menlo Systems, FC1500-250-WG) with an $f_{rep}$ of 250 MHz. The comb spectrum is shifted to cover 6 THz around 1.68 μm using a polarization-maintaining microstructured silica fiber[37]. The comb is locked to the cavity using the two-point PDH stabilization scheme[23], where two error signals are derived from the light reflected from the cavity, picked up using a fiber optical circulator and dispersed by a free-space reflection grating. Two selected ranges of the dispersed light, referred to as locking points, are incident on two high-bandwidth photodetectors. Correction signals are derived from the two detectors using proportional-integral controllers and sent to the current of the oscillator pump diode, which controls $f_{rep}$ and $f_{ceo}$, and to a PZT and electro-optic modulator in the oscillator cavity that control the $f_{rep}$. Absolute frequency stability is ensured by locking $f_{rep}$ to the output of a tunable direct digital synthesizer (DDS) referenced to a GPS-disciplined Rb oscillator via actuating on the sample cavity length, similar to what was done in Ref. [26]. The $f_{ceo}$ is indirectly stabilized via the comb-cavity lock and monitored using an $f$-$2f$ interferometer during the acquisition of the spectra.

The 80-cm-long cavity is made of two mirrors (Layertec) with 5-m radius of curvature, with maximum reflectivity and minimum dispersion at the design wavelength of 1580 nm. The mirror transmission at the pump wavelength is 60%. The transmitted comb has a mode spacing of 750 MHz, resulting from the 4:3 ratio of the incident comb $f_{rep}$ and the cavity *FSR*. We note that this mode filtering is not necessary for the operation of the technique, but it is a result of using a cavity from a different experiment[38], where such filtering was needed. The comb beam is mode matched to the $TEM_{00}$ transverse mode of the cavity, which has a Rayleigh range of 1.4 m at 1650 nm. To maximize the spatial overlap between the pump and probe beams in the cavity, the pump beam is mode matched to have its waist in the middle of the cavity and the same Rayleigh range, which corresponds to a pump beam waist of 1.2 mm at 3.3 μm. The pump power incident on the sample, calculated as the power transmitted through the cavity when the pump is off-resonance, divided by the mirror transmission, is 180 mW, 165 mW, and 150 mW for the P(2, $F_2$), Q(2, $F_2$) and R(2, $F_2$)-pumped spectra, respectively, and the fractional pump transmission on resonance (*i.e.*, when locked to the center of the pump transition) is 64%, 70% and 61%, respectively.

***Spectral acquisition***

The probe comb spectra are measured using a home-built Fourier transform spectrometer (FTS) with auto-balanced detection, previously used in Refs. [21,22,26]. The optical path difference is calibrated using a stabilized 633-nm HeNe laser (Sios, SL/02/1), which has a fractional frequency stability of $5 \times 10^{-9}$ over 1 h. The comb interferograms are recorded simultaneously with the reference laser interferograms by a digital oscilloscope (National Instruments, PCI-5922) with a sampling rate of 5 MS/s and a 20-bit resolution. The comb interferograms are interpolated at the zero-crossings and extrema of the corresponding HeNe interferogram.

During the acquisition, the sample and background interferograms are measured with the pump unblocked and blocked on consecutive FTS scans using the shutter (Thorlabs, SHB1T). To record the narrow sub-Doppler transitions, the sample point spacing needs to be much smaller than the $f_{rep}$. Therefore, the comb $f_{rep}$ – and with it the cavity *FSR* – are tuned by stepping the DDS frequency, and spectra are recorded at 387 values of $f_{rep}$ differing by 2.75 Hz, which corresponds to a shift of comb modes by 2 MHz in the optical domain. For each step, the nominal resolution of the spectrometer is matched in postprocessing to the $f_{rep}$, and the frequency scale is shifted by the $f_{ceo}$, which yields spectra with comb-calibrated frequency axis[25,26].

The acquisition time of one interferogram with nominal resolution of 750 MHz is 1.3 s, which yields a total acquisition time of 16.7 min for one normalized and interleaved spectrum, given by 1.3 × 2 × 387 s, where the factor of 2 comes from the acquisition of spectra with and without the pump at each step.

### *Comparison to TheoReTS/HITEMP*

All probe transitions measured with the pump locked to the $\nu_3$ P(2, $F_2$) and $\nu_3$ Q(2, $F_2$) transitions, and 4 measured with the pump locked to the $\nu_3$ R(2, $F_2$) transition, could be unambiguously matched to the closest strongest TheoReTS line, with the assignment of the final state agreeing with the experimental result. When more than one strong TheoReTS line was within the claimed TheoReTS accuracy of 1 $cm^{-1}$ from the measured probe transition, we used the ratios of experimental and TheoReTS integrated absorptions to verify the match. Ideally, one would use the intensities of the so-called V-type transitions in the $2\nu_3$ band transitions to estimate the fraction of the population transferred to the upper pump level as was done in Ref. [22]. However, V-type transitions are not observed in the present cavity-enhanced spectra, because the absorption of the $2\nu_3$ band transitions is fully saturated at 50 mTorr, and there were no other transitions with lower state rotational quantum number $J = 2$ within the probed spectral range. Therefore, we made a relative comparison of the integrated absorption of the sub-Doppler probe lines to the integrated absorption of the Doppler-broadened TheoReTS lines. We calculated the latter as a product of the TheoReTS line intensity and the sample density at 50 mTorr and 296 K, which is $1.63 \times 10^{15}$ molecules $cm^{-3}$. The experimental and predicted integrated absorptions are listed in Supplementary Table S2 in the Supplementary information, while their ratios are shown in Fig. 6(b). For the unambiguously matched lines in the P(2, $F_2$), Q(2, $F_2$) and R(2, $F_2$)-pumped spectra, the mean ratios are 5900(700), 6200(400), and 6100(1100), respectively. The absolute values of these ratios reflect the difference in the population of the upper pump level in the OODR experiment compared to thermal population at room temperature. We matched another 8 lines in the R(2, $F_2$)-pumped spectrum by choosing the TheoReTS line for which the ratio of the experimental and the predicted intensities was closest to the mean for the unambiguously assigned lines. The mean intensity ratio of the lines in the R(2, $F_2$)-pumped spectrum, neglecting the two outliers visible in Fig. 6(b), is 6400(1200).

**Data availability statement**: The data that support the findings of this study are available from the corresponding author upon reasonable request.

**Acknowledgments**: The authors thank Hiroyuki Sasada for providing the ground state term value from his unpublished work, and Michael Rey for providing the final state assignments from the non-empirical effective Hamiltonian described in his work, Ref. [33].

**Author contributions**: K.K.L. and A.F. conceived the idea. V.S.O. and I.S. implemented the experiment and performed the measurements. V.S.O analyzed and visualized the data. L.R. and K.K.L. contributed theoretical predictions and analysis tools. A.F., G.S., and O.A. provided resources. A.F. supervised the project and wrote the manuscript. V.S.O, L.R., G.S., O.A, and K.K.L. revised the manuscript.

**Competing interest statement**: The authors declare no competing interests.

**Funding**: A.F. acknowledges the Knut and Alice Wallenberg Foundation (KAW 2015.0159, KAW 2020.0303) and the Swedish Research Council (2020-00238). L.R. acknowledges the French National Research Agency (ANR-19-CE30-0038-01); G.S. acknowledges the Foundation for Polish Science (POIR.04.04.00-00-434D/17-00); O.A. acknowledges the Kempe Foundation (JCK 1317.1), K.K.L. acknowledges the National Science Foundation (grant: CHE- 2108458).

# Sub-Doppler optical-optical double-resonance spectroscopy using a cavity-enhanced frequency comb probe: Supplementary information

**Vinicius Silva de Oliveira[1], Isak Silander[1], Lucile Rutkowski[2], Grzegorz Soboń[3], Ove Axner[1], Kevin K. Lehmann[4], and Aleksandra Foltynowicz[1,*]**

[1] Department of Physics, Umeå University, 901 87 Umeå, Sweden

[2] Univ Rennes, CNRS, IPR (Institut de Physique de Rennes)-UMR 6251, F-35000 Rennes, France

[3] Faculty of Electronics, Photonics and Microsystems, Wrocław University of Science and Technology, Wybrzeże Wyspiańskiego 27, 50-370 Wrocław, Poland

[4] Departments of Chemistry & Physics, University of Virginia, Charlottesville, VA 22904, USA

Corresponding author: aleksandra.foltynowicz@umu.se

## 1. Cavity finesse

The cavity finesse was evaluated from a measurement of the cavity ring-down time at 11 points between 5910 and 5980 $cm^{-1}$. The mean of 100 retrieved finesse values at each wavenumber is shown by the black markers in Fig. S1, where the error bars are standard deviations of the mean. A linear fit to the finesse (red curve) was used to fix the finesse values in the model for line fitting. The design mirror wavelength (with maximum reflectivity) was 1580 nm (6300 $cm^{-1}$), and the linear fit agrees well with the mirror reflectivity data. The 1σ confidence interval of the fit (shaded grey) indicates 3% relative uncertainty of the finesse, which was propagated to the retrieved integrated absorption values.

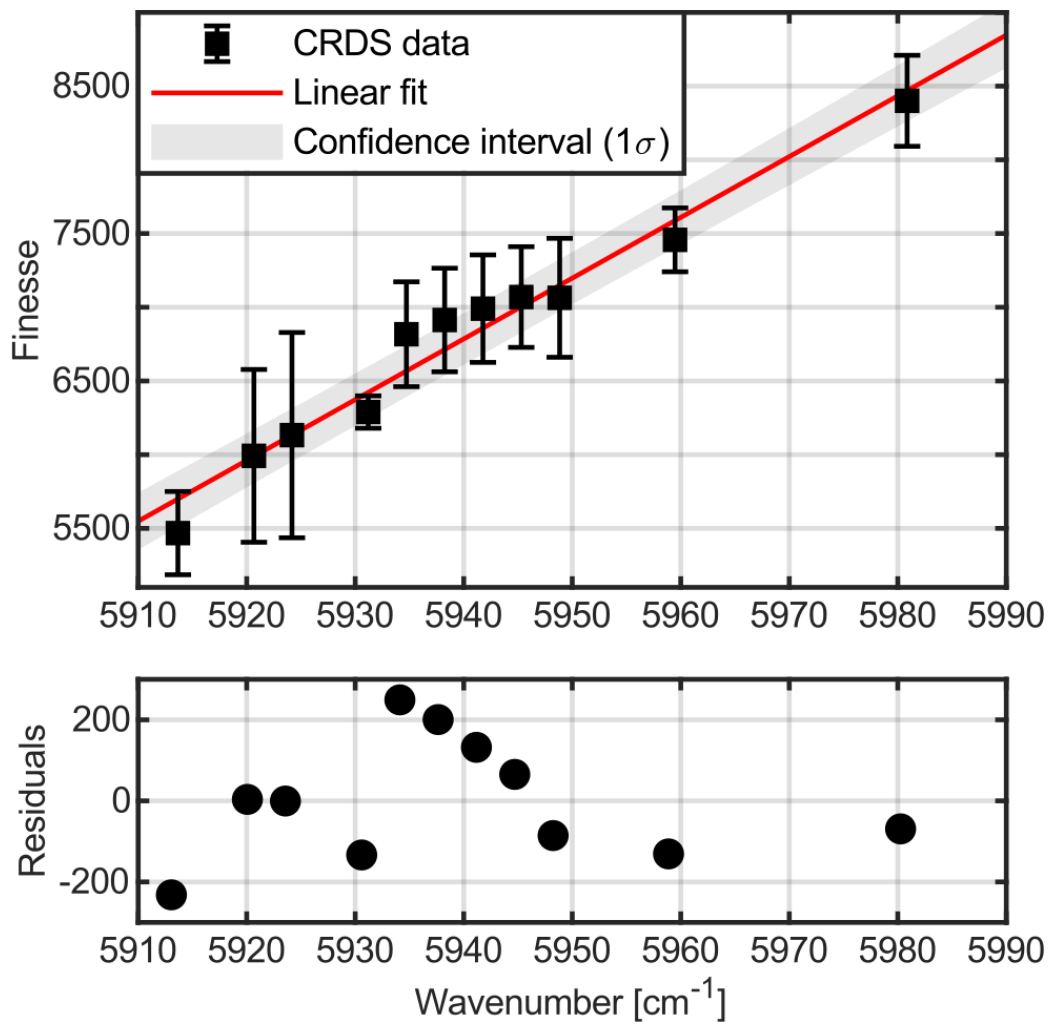

Figure S1. **Cavity finesse.** The cavity finesse evaluated from the cavity ring-down time (black markers) together with a linear fit to the data (red curve), the 1σ confidence interval (shaded) and the fit residuals (lower window).

## 2. Data averaging

During the acquisition, the stepping of the repetition rate was performed in two different ways. For measurements with the pump locked to the R(2, $F_2$) transition, 5 pairs of sample and background interferograms were taken at each $f_{rep}$ value. The $f_{rep}$ was then stepped using the procedure described in the main paper. For the parallel relative pump/probe polarizations, this process was repeated 9 times to acquire a total of 45 interferograms at each $f_{rep}$ value. For measurements with the pump locked to the P(2, $F_2$) and Q(2, $F_2$) transitions, one pair of sample and background interferograms was measured, then the $f_{rep}$ was stepped, and a new pair was measured. This scan was repeated 5 times to yield 5 averages. The P(2, $F_2$)-pumped dataset was measured twice, with different Pound-Drever-Hall (PDH) comb-cavity locking points, to optimize the SNR in the spectral region where probe transitions are found.

The $f_{\text{ceo}}$ was indirectly stabilized via the two-point PDH lock[1] and monitored using a counter with 1 s integration time. We found that during the long $f_{\text{rep}}$ scans the $f_{\text{ceo}}$ drifted by up to 200 kHz, which we attribute to drifts in the offsets of the PDH locks. This implied that the frequency axes in the five consecutive P(2, $F_2$)- and Q(2, $F_2$)-pumped spectra were known accurately but were not identical. Therefore, instead of averaging the 5 spectra, we combined them into one interleaved spectrum (*i.e.*, having 5 points at each $f_{\text{rep}}$ value). The measurement with the pump locked to the R(2, $F_2$) transition could be averaged 5 times, since we acquired 5 consecutive sample-background pairs at each $f_{\text{rep}}$ step, during which the $f_{\text{ceo}}$ was constant. However, for consistency, for the fitting, we combined the R(2, $F_2$)-pumped spectra in the same way as the P(2, $F_2$)- and Q(2, $F_2$)-pumped spectra. The weakest lines in the R(2, $F_2$)-pumped spectrum could be observed only after averaging all 45 spectra from the 9 scans of $f_{\text{rep}}$. To average the spectra from the different $f_{\text{rep}}$ scans that had slightly different $f_{\text{ceo}}$ values, we interpolated the data to the same frequency axis, *i.e.*, the same $f_{ceo}$ value.

## 3. Line fitting

To retrieve the parameters of probe transitions, we fit the cavity transmission model[1], where the line shape was assumed to comprise a sum of a narrow and strong Lorentzian function for the sub-Doppler probe transition, and a wider and weaker Gaussian function for the Doppler-broadened background absorption originating from thermal redistribution of the population of the upper pump level by elastic velocity changing collisions[2]. The fitting range was ±500 MHz, which is more than 3 times the width of the Doppler-broadened background. Figure S2 shows examples of fits (zoomed to ±250 MHz for clarity) to two probe lines in the R(2, $F_2$)-pumped spectrum with each type of contribution indicated separately. The fit model was implemented in Matlab and fitted using the trust-region nonlinear least squares algorithm. The fit parameters were the integrated absorptions and widths of the Lorentzian and Gaussian profiles, the center frequency common for the Lorentzian and Gaussian profiles, and the wavelength-dependent comb-cavity phase offset. Dispersion in the cavity mirror coatings causes a non-zero comb-cavity resonance offset away from the PDH locking points. The cavity-transmission function also contains molecular dispersion, which results in an additional shift of the cavity resonances with respect to the comb lines. This shift has a dispersive shape as a comb mode is moved across an absorption line.[1] Since this molecular contribution to the shift is small compared to the width of the cavity modes, the change in cavity transmission is approximately linear with this molecular induced shift. This effect is well understood and included in the model, and it does not shift the fitted center frequency. The cavity finesse was fixed to the values from the fit to the experimental data (see Section 1). For the weak lines, the Gaussian width was fixed to 141.7 MHz, while the ratio of the sub-Doppler and Doppler-broadened integrated absorptions was fixed to 0.72, based on the mean of the values from fits to strong lines. The sidebands originating from the frequency modulation used in the Lamb dip stabilization of the pump that appear at ±120 MHz around the central peak (twice the modulation frequency due to the probe being twice the pump frequency) were excluded from the fit and are marked in gray in Fig. S2. For probe lines overlapping with Doppler-broadened transitions from the $2\nu_3$ band, the absorption and dispersion of the $2\nu_3$ band lines (based on the parameters from the HITRAN database[3]) were included in the model of the background spectrum, as their presence modifies the cavity losses, and thus the enhancement, even though they are removed by normalization.

The structure visible in the residuals around the line centers indicates that the model based on a sum of a single Lorentzian and single Gaussian function is not fully sufficient to describe the observed line shape. Attempts to include a sum of individual Lorentzian functions for the different $M_J$ components, where $M_J$ is the quantum number for the projection of total angular momentum on the axis defined by the pump electric field, did not yield an improvement. More accurate modeling of the observed line shape will be a subject of further study. For now, we note that the residuals are symmetric around the line center, and thus the inaccuracy in the model does not affect the accuracy of the center frequency retrieval.

We note that in Fig. S2 all measurement points are combined (interleaved) as described in Section 2, whereas in Fig. 2 in the main paper the 5 spectra are averaged for clarity.

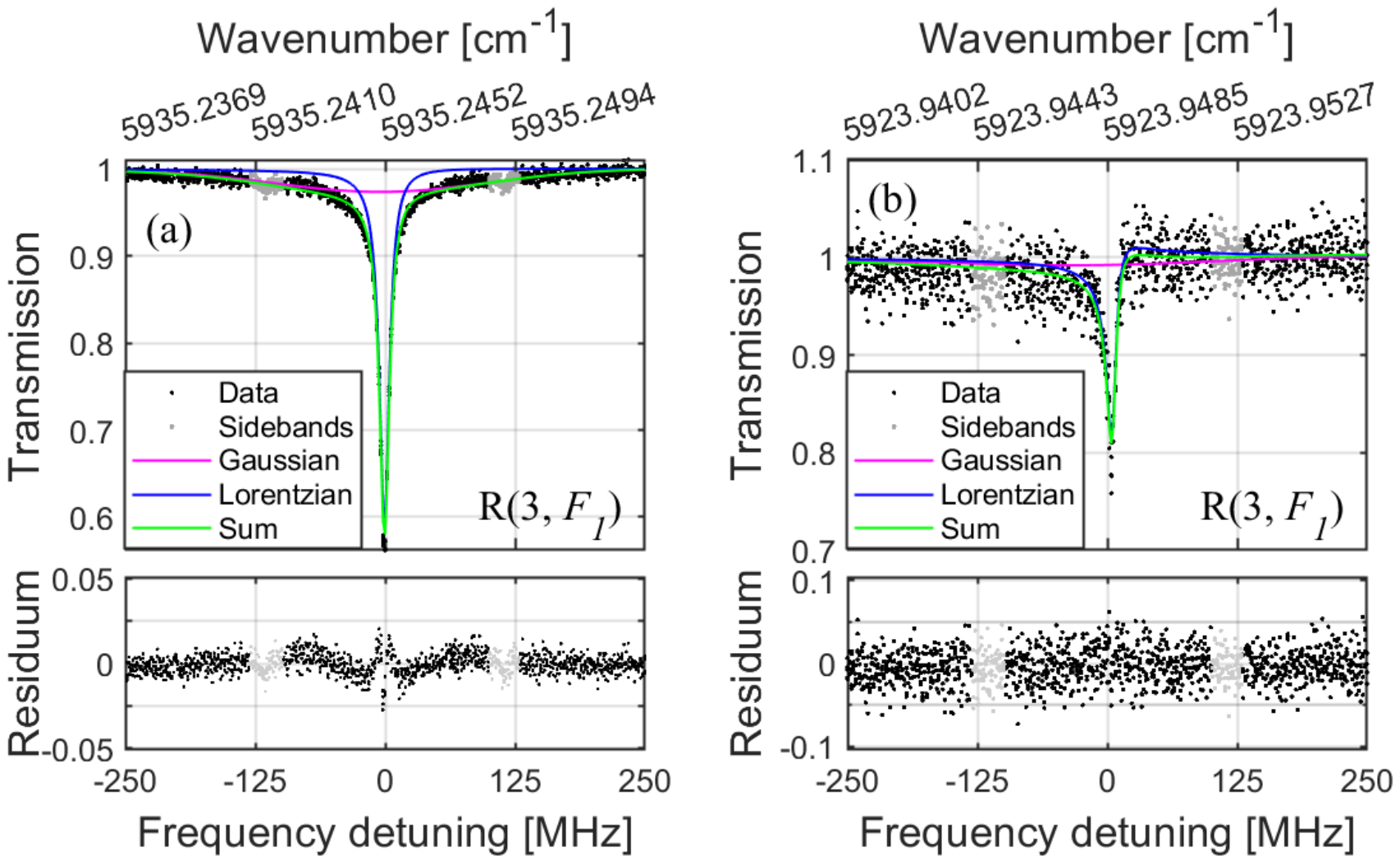

Figure S2: **Fits to probe lines with high and low SNR.** Comparison of fits to probe lines with (a) high SNR and (b) low SNR detected in the spectrum with the pump locked to the $\nu_3$ R(2, $F_2$) transition with parallel relative pump/probe polarizations. The 5 combined spectra are shown in black in the upper windows. The green curves show the fit of the model, while the blue and magenta curves display the Lorentzian and Gaussian parts, respectively. The lower windows show residuals of the fits. The ranges around the FM sidebands that are excluded from fitting are indicated in gray.

## 4. Frequency uncertainty

### *4.1 Frequency scale calibration*

The frequency scale was calibrated using the sub-nominal resolution method of Refs. [4,5], where precise sampling of the comb modes is obtained by setting the nominal resolution of the Fourier transform spectrometer (FTS) equal to comb mode spacing and shifting the origin of the FTS frequency axis to account for the $f_{ceo}$. The fine calibration of the frequency axis was performed by minimizing the instrumental line shape on the strongest sub-Doppler probe line in each spectrum. To do that, we generated spectra of the strongest line using a set of values of the reference laser wavelength (which calibrates the optical path difference, and thus the nominal resolution) differing by ~10 fm. Afterward, we performed fits to the resulting spectra, as described in Section 3, and plotted the standard deviation of the fit residuals as a function of the reference laser wavelength. The minimum of the standard deviation corresponds to the lowest instrumental line shape and thus to the best match between the comb mode spacing and the FTS nominal resolution. From the depth of this minimum, we estimated the uncertainty in the reference laser wavelength to be 50 fm. Finally, we fit a line to the center frequencies from fits to those spectra plotted as a function of the reference laser wavelength. From the slope of this line and the 50 fm uncertainty of the reference laser wavelength, we obtained an uncertainty of the center frequency of 30 kHz (1 $\times 10^{-6}$ cm$^{-1}$), which is comparable to the fit uncertainty for the strongest lines, but negligible compared to the total uncertainty (see Section Frequency uncertainty in the main paper and 4.2 below).

We note that in Ref. [5] it was shown that when the repetition rate of the comb is much larger than the line width of the transition (as is the case here for the sub-Doppler probe lines) the center frequency should not change with reference laser wavelength calibration. However, here, the sub-Doppler lines reside on top of a Doppler-broadened component, whose width is of the same order as $f_{\mathrm{rep}}$. In the fit, the center frequency is assumed equal for both components, therefore, the center frequency depends slightly on the reference laser wavelength calibration.

### *4.2 Frequency uncertainty model*

As discussed in the main paper, to investigate the long-term stability of the center frequency, we performed fits to 9 probe transitions detected in the 45 spectra recorded over 12 h with the pump locked to the $\nu_3$ R(2, $F_2$) transition and parallel relative pump/probe polarization. The analysis of the center frequencies from the 45 individual fits to each line revealed that their spread is larger than the uncertainty of the individual fits, which we attribute to a residual uncorrected baseline drift.

Figure S3 shows the 1σ standard deviation of the center frequencies from the long-term measurement (green markers, left axis) and the fit precision of all probe lines detected in the five combined P(2, $F_2$), Q(2, $F_2$), and R(2, $F_2$)-pumped spectra (blue, red, and black markers, respectively), as a function of the SNR of the line in a single measurement. The SNR was calculated as the ratio of the peak absorption (1-transmission) and the standard deviation of the residuals (excluding the FM sidebands and the line center due to the mismatch between the model and the measurement observed for high-SNR lines). For the long-term series (green markers), the SNR was taken as the mean SNR of the 45 consecutive spectra, where the horizontal error bar is the standard deviation of the mean. For the probe lines fitted in the five combined spectra (blue, red, and black markers, respectively, right axis), the standard deviation of the residuals corresponds to the noise in a single measurement rather than an average of 5 measurements, since the data are interleaved rather than averaged (as described in Section 2).

Since the uncertainty from the long-term measurement is systematically larger than the fit precision for all lines, we developed a model for the frequency uncertainty based on the SNR dependence of the standard deviation of the long-term measurements. The blue curve in Fig. S3 shows a fit of a model function $U = (a^2+b^2/\mathrm{SNR}^2)^{1/2}$ to the observed relation between the uncertainty and the SNR from the 45-measurement series, where $a$ and $b$ are fitting parameters. We used this function to estimate the center frequency uncertainty for all detected lines based on their single-measurement SNR evaluated as described above. Since the baseline fluctuations, to which we attribute the increased frequency uncertainty, do not depend on the choice of the pump transition, we applied the model also to the lines that were not measured 45 times, *i.e.*, the P(2, $F_2$)- and Q(2, $F_2$)-pumped lines, and the 6 weakest R(2, $F_2$)-pumped lines visible only in the spectrum averaged 45 times.

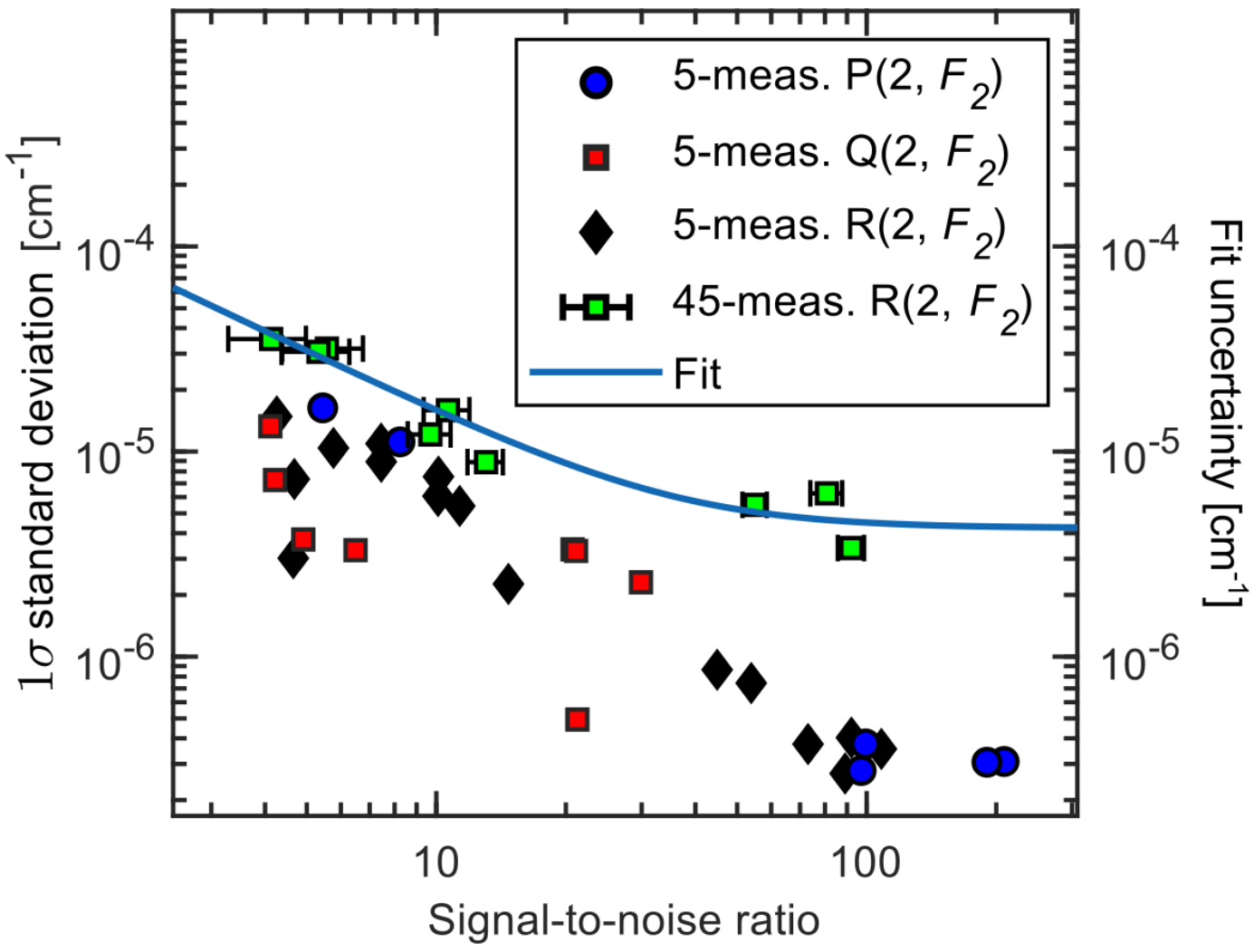

Figure S3. **Frequency uncertainty model.** Center frequency uncertainty as a function of the SNR of the individual lines in a single measurement. The green markers (left vertical axis) show the standard deviations of the center wavenumbers evaluated from fits to the long-term measurements with the pump locked to the R(2, $F_2$) transition. The blue, red, and black markers (right vertical axis) show center frequency fit precision for all lines detected with the pump on the P(2, $F_2$), Q(2, $F_2$), and R(2, $F_2$) transitions, respectively.

## 5. Comparison to previous measurements

Two probe transitions in the spectrum measured with the pump locked to the $\nu_3$ P(2, $F_2$) transition were observed in our previous work[2], where the sample was contained in the liquid-nitrogen-cooled single-pass cell. Figure S4 shows one of these transitions measured in the cell (from Ref. [2], 16 averages, 40 mTorr, 110 K, measurement with pump beam on only) and in the cavity (this work, 1 average, 296 K, 50 mTorr, ratio of spectra with pump on and off). The SNR in the cell was 10 after 3.2 h of averaging, while in the cavity it is 200 in 16.7 minutes, *i.e.*, less than 1/10 of the time. Normalized to the same acquisition time, this implies that the SNR in the cavity is a factor of $200 \times 10^{-1/2} = 60$ better than in the cell. This improvement is smaller than the factor of 700 improvement in the noise equivalent absorption sensitivity (see Section Sensitivity in the main manuscript) because the pump absorption was stronger in the cooled cell. The temperature dependence of the absorption coefficient of the pump transitions at a constant pressure is given by $(T_2/T_1)^{-3} \times \exp[-(c_2E_{rot})(1/T_2 - 1/T_1)]$, where the $T^{-3}$ dependence is a product of the temperature dependence of the number density ($T^{-1}$), the inverse of the partition function ($T^{-3/2}$ for a nonlinear molecule), and the normalized line shape function ($T^{-1/2}$ for Doppler-broadened transitions), and the second term represents the ratio of the Boltzmann factors, where $E_{rot}$ is the lower pump level term value and $c_2$ is the second radiation constant. For $T_1 = 296$ K, $T_2 = 110$ K, and $E_{rot} = 31.4423878$ cm$^{-1}$, this yields a factor of 14.6. Considering also the difference in sample pressure (50 mTorr vs 30 mTorr), the absorption coefficient of the pump transitions was 11.7 stronger in the cell than in the cavity. This implies that the improvement in the SNR can be estimated to 700/11.7 = 60, which agrees with what is observed in the data. We note that in the main paper we state the improvement of 700/14.6 = 50, which is valid under same pressure conditions.

A comparison of the wavenumbers and upper state term values of the two lines detected both in the cell and in the cavity is shown in Table S1, demonstrating a more than 1 order of magnitude improvement in frequency accuracy. This improvement is confirmed by the fact that all final states in the combination differences agree within their uncertainties.

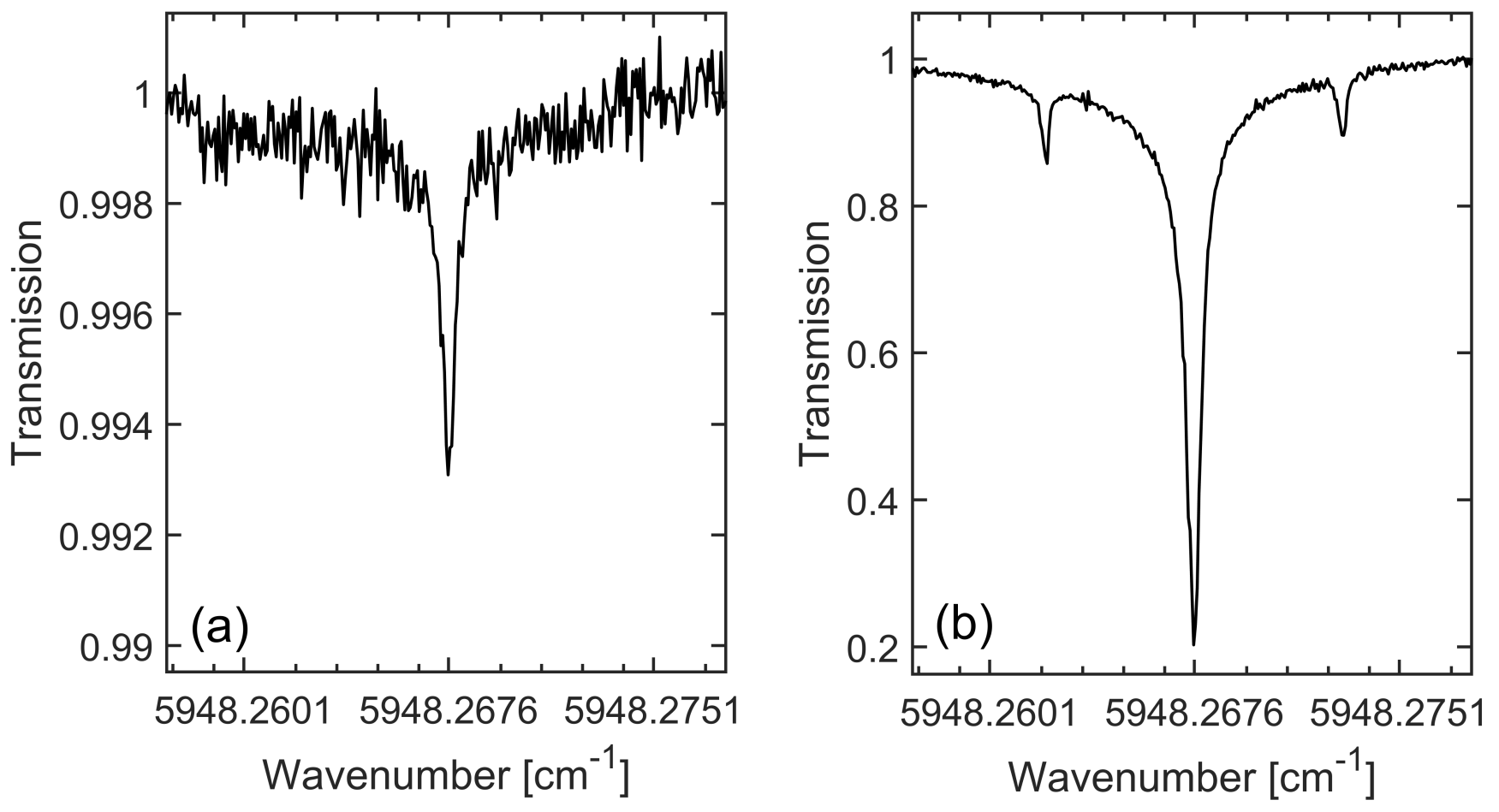

Figure S4. **Sensitivity comparison.** The R(1) probe line detected in the spectrum measured with the pump locked to the ν$_3$ P(2, $F_2$) transition in (a) the single pass cell (from Ref. [2]) and (b) in the enhancement cavity in this work.

Table S1: **Comparison between probe transition wavenumbers and final state term values** for two OODR probe lines detected both in the cell[2] and in the cavity (this work).

| Pump transition, $v_3$ band | Probe transition | Probe transition wavenumber [cm$^{-1}$] | Final term state value [cm$^{-1}$] | Ref. |
|---|---|---|---|---|
| P(2, $F_2$) | R(1) | 5979.04297(5) | 9009.47939(5) | [2] |
| | | 5979.042972(4) | 9009.479391(4) | This work |
| | | | | |
| P(2, $F_2$) | R(1) | 5948.26760(5) | 8979.70402(5) | [2] |
| | | 5948.267590(3) | 8979.704010(3) | This work |

## 6. Predicted and experimental polarization-dependent intensity ratios

As explained in Ref. [6], the polarization-dependent intensity ratios can be predicted as the ratio of the sum over $M_J$ quantum states of the products of the absolute value of the transition dipole for the pump transition times the square of the transition dipole for the probe transition evaluated for each relative polarization. The reason this ratio is proportional to the pump transition dipole rather than its square is that in the strong pumping limit, the steady-state population integrated over detuning is proportional to the Rabi frequency. The linear polarization-dependent intensity ratios predicted for strongly saturated inhomogeneously-broadened pump transition and unsaturated probe transition are given in Table IV of Ref. [6]. The predicted intensity ratios for all pump-probe combinations observed in this work are listed in the second last column of Table II in the main manuscript.

The accuracy of the measured polarization-dependent intensity ratios was limited by the quality of the probe polarization. This polarization was slightly elliptical because of the propagation in the nonlinear fiber and the fiberized optical circulator. 4.2% of the probe power in front of the cavity was along the minor axis, which corresponds to an ellipticity of 0.2. Compared to that, the pump polarization was purely linear. To calculate the uncertainties in the measured intensities and their ratios caused by the ellipticity of the probe polarization, we assumed that the polarizations of the pump and probe are linear but offset by an angle $\varphi$ = 0.2 from being perfectly parallel or perpendicular.

For the integrated intensities measured with parallel and perpendicular relative pump/probe polarizations, $I_{\parallel}$ and $I_{\perp}$, we assumed a relative error of $sin^2(\varphi)$ = 4.2% caused by the ellipticity of the probe light.

To evaluate the uncertainty in the measured polarization-dependent intensity ratios, we compared the intensity ratios predicted for perfectly parallel and perpendicular pump and probe polarizations, given by

$$R_{\parallel/\perp}(0) = \frac{I_{\parallel}}{I_{\perp}}, \tag{1}$$

with ratios predicted for an angle $\varphi$, given by

$$R_{\parallel/\perp}(\varphi) = \frac{I_{\perp} sin^2(\varphi) + I_{\parallel} cos^2(\varphi)}{I_{\perp} cos^2(\varphi) + I_{\parallel} sin^2(\varphi)} = \frac{R_{\parallel/\perp}(0) + sin^2(\varphi)\left[1 - R_{\parallel/\perp}(0)\right]}{1 - sin^2(\varphi)\left[1 - R_{\parallel/\perp}(0)\right]}. \tag{2}$$

We then took twice the difference $2\left|R_{\parallel/\perp}(\varphi) - R_{\parallel/\perp}(0)\right|$, given by

$$2\left|R_{\parallel/\perp}(\varphi) - R_{\parallel/\perp}(0)\right| = \frac{2 sin^2(\varphi)\left[1 - R^2_{\parallel/\perp}(0)\right]}{1 - sin^2(\varphi)\left[1 - R_{\parallel/\perp}(0)\right]}, \tag{3}$$

as the uncertainty in the measured polarization-dependent intensity ratio introduced by the probe ellipticity $\varphi$. We note that this uncertainty is smallest for ratios closest to 1, and larger for ratios that are farther from 1.

## 7. Results for all probe lines

Table S2 contains the results of the fits to all detected probe lines. Column 1 shows the pump transition in the $\nu_3$ band. Column 2 lists the experimental probe transition assignment (in agreement with TheoReTS/HITEMP[7]). The 6 weakest transitions, marked by an asterisk, were detected only in the spectrum averaged 45 times with parallel pump/probe polarizations, and their assignments (if available) are from TheoReTS/HITEMP only. Column 3 lists the measured center wavenumber, calculated as a weighted mean of the wavenumbers found from fits to spectra recorded with the two pump polarizations, where the weight was the inverse of the square of the uncertainty calculated from the SNR-based model (see Section 4.2). Column 4 shows the final state term value, calculated as described in Section Line assignments in the main paper. Column 5 states the integrated absorption of the probe transition, calculated as a weighted mean of the absorption measured with parallel and perpendicular relative pump/probe polarizations, with weight 1 for parallel and 2 for perpendicular polarization, which yields an isotropic average independent of the value of the $M_J$ quantum number. For the 6 weakest lines that were measured only with parallel pump/probe polarization, we recalculated the measured absorption to the weighted mean using the predicted polarization-dependent intensity ratios. The uncertainty is a combination of the fit uncertainty, the uncertainty in the finesse determination (3%), and polarization of the probe (4.2%). Column 6 provides the ratio of the line intensities measured with the parallel and perpendicular relative pump/probe polarizations. The uncertainty is a combination of the fit uncertainty of the two integrated absorptions, and the uncertainty caused by the ellipticity of the probe given by Eq. (3) above. The uncertainty in the finesse is not taken into account, since the contribution from the finesse cancels when taking the ratio of the two intensities. Column 7 shows the half width at half maximum of the probe transition, calculated as a weighted mean of the widths found from fits to spectra recorded with the two pump polarizations, where the weight was the inverse of the fit variance. Columns 8 and 9 display the integrated absorption and the width of the Gaussian part, respectively. For the weaker lines, the ratio of the sub-Doppler and the Doppler-broadened integrated absorptions was fixed to 0.72, i.e. the mean value found from the fits to the strongest lines. Similarly, the width of the Doppler-broadened contribution was fixed to 141.7 MHz, and it is given without uncertainty in column 9. This width is slightly smaller than the 165 MHz thermal Doppler width at 296 K. Columns 10 and 11 show the predicted probe transition wavenumber from TheoReTS/HITEMP[7] and its difference with respect to the observed wavenumber [plotted in Fig. 6(c) in the main paper]. Column 12 shows the final state dominant assignment obtained from the Hamiltonian described in Ref. [8]. Columns 13 and 14 give the line intensity from TheoReTS/HITEMP at 296 K[7] and the integrated absorption calculated at the experimental conditions of 296 K and 50 mTorr.

Table S2: **Results of fits to all lines.** See description in text.

| 1 | 2 | 3 | 4 | 5 | 6 | 7 | 8 | 9 | 10 | 11 | 12 | 13 | 14 |
|---|---|---|---|---|---|---|---|---|---|---|---|---|---|
| Pump transition in the $\nu_3$ band | Probe transition | Probe transition wavenumber [cm$^{-1}$] | Final state term value [cm$^{-1}$] | Probe transition integrated absorption [$10^{-9}$ cm$^{-2}$] | Polar. depend. intensity ratio | Probe transition width [MHz] | Gaussian integrated absorption [$10^{-9}$ cm$^{-2}$] | Gaussian width [MHz] | TheoReTS transition wavenumber [cm$^{-1}$] | Obs. – pred. wavenum. [cm$^{-1}$] | Final state assignment from Ref. [8] | TheoReTS transition intensity [$10^{-30}$ cm/molec] | TheoReTS integrated absorption [$10^{-13}$ cm$^{-2}$] |
| P(2,$F_2$) | R(1) | 5948.267590(3) | 8978.704010(3) | 2.26(9) | 0.91(5) | 5.093(8) | 1.50(3) | 144.3(4) | 5947.67695 | 0.59 | $3\nu_3$ ($F_1$) | 220.40 | 3.59 |
| | R(1) | 5964.06227(2) | 8994.49869(2) | 0.081(4) | 1.01(8) | 6.5(3) | 0.058(2) | 141.7 | 5965.37197 | -1.31 | $5\nu_2+3\nu_4$ ($F_1$) | 7.79 | 0.13 |
| | R(1) | 5979.042972(3) | 9009.479392(3) | 0.56(2) | 0.98(6) | 4.459(8) | 0.396(9) | 125.5(6) | 5979.71721 | -0.67 | $\nu_1+4\nu_2$ ($A_1$) | 68.70 | 1.12 |
| Q(2,$F_2$) | Q(2,$F_1$) | 5928.61142(2) | 8978.70401(2) | 0.56(4) | 1.5(3) | 6.0(1) | 0.40(2) | 141.7 | 5928.02556 | 0.59 | $3\nu_3$ ($F_1$) | 53.28 | 0.87 |
| | Q(2,$F_1$) | 5944.40608(2) | 8994.49868(2) | 0.103(5) | 1.5(3) | 4.31(9) | 0.074(2) | 141.7 | 5945.72057 | -1.31 | $5\nu_2+3\nu_4$ ($F_1$) | 10.83 | 0.18 |
| | R(2,$F_1$) | 5958.673574(6) | 9008.766169(6) | 2.38(10) | 0.83(7) | 4.73(8) | 2.20(8) | 143(4) | 5957.83620 | 0.84 | $3\nu_3$ ($F_1$) | 243.60 | 3.97 |
| | Q(2,$F_1$) | 5959.386797(5) | 9009.479392(5) | 0.89(3) | 1.8(3) | 4.29(1) | 0.64(1) | 141.7 | 5960.06582 | -0.68 | $\nu_1+4\nu_2$ ($A_1$) | 82.82 | 1.35 |
| R(2,$F_2$) | R(3,$F_1$) | 5913.18732(2) | 8992.78303(2) | 0.062(4) | 1.1(1) | 3.6(3) | 0.045(3) | 141.7 | 5913.97472 | -0.79 | $3\nu_2+\nu_3+\nu_4$ ($F_2$) | 6.48 | 0.11 |
| | R(3,$F_1$) | 5918.14141(1) | 8997.73712(1) | 0.096(5) | 1.05(10) | 4.1(1) | 0.069(3) | 141.7 | 5918.79890 | -0.66 | $3\nu_2+\nu_3+\nu_4$ ($F_2$) | 11.14 | 0.18 |
| | *R(3,$F_1$) | 5920.775507(9) | 9000.371213(9) | 0.039(2) | == | 5.67(6) | 0.034(1) | 141.7 | 5921.56979 | -0.79 | $3\nu_2+\nu_3+\nu_4$ ($F_2$) | 4.43 | 0.07 |
| | * == | 5922.51469(4) | 9002.11039(4) | 0.012(2) | == | 4(1) | 0.010(2) | 141.7 | == | == | | == | == |
| | *R(3,$F_1$) | 5923.32503(5) | 9002.92074(5) | 0.039(6) | == | 9(2) | 0.033(4) | 141.7 | 5923.25121 | 0.074 | == | 2.17 | 0.04 |
| | R(3,$F_1$) | 5923.94848(1) | 9003.54418(1) | 0.44(2) | 1.19(10) | 5.2(2) | 0.32(1) | 141.7 | 5924.39254 | -0.44 | $3\nu_2+\nu_3+\nu_4$ ($F_1$) | 49.03 | 0.80 |
| | R(3,$F_1$) | 5924.26536(2) | 9003.86107(2) | 0.175(7) | 1.1(1) | 3.79(8) | 0.126(3) | 141.7 | 5924.10232 | 0.16 | $3\nu_2+\nu_3+\nu_4$ ($F_2$) | 13.00 | 0.21 |
| | *R(3,$F_1$) | 5924.99660(1) | 9004.59230(1) | 0.049(4) | == | 4.3(3) | 0.041(2) | 141.7 | 5924.99972 | -0.003 | $3\nu_2+\nu_3+\nu_4$ ($F_2$) | 2.23 | 0.04 |
| | * == | 5929.14547(1) | 9008.74117(1) | 0.112(3) | == | 5.67(9) | 0.036(1) | 141.7 | == | == | | == | == |
| | Q(3,$F_1$) | 5929.170466(3) | 9008.766172(3) | 4.7(2) | 0.4(1) | 5.87(1) | 2.81(7) | 142.4(3) | 5928.34007 | 0.83 | $3\nu_3$ ($F_1$) | 418.90 | 6.83 |
| | P(3,$F_1$) | 5929.883687(4) | 9009.479393(4) | 0.85(3) | 1.6(2) | 4.51(2) | 0.61(1) | 141.7 | 5930.56969 | -0.69 | $\nu_1+4\nu_2$ ($A_1$) | 71.45 | 1.16 |
| | R(3,$F_1$) | 5932.279186(9) | 9011.874892(9) | 0.100(4) | 1.2(1) | 4.00(6) | 0.072(2) | 141.7 | 5932.52779 | -0.25 | $3\nu_2+\nu_3+\nu_4$ ($F_2$) | 8.00 | 0.13 |
| | *R(3,$F_1$) | 5932.876751(8) | 9012.472457(8) | 0.025(2) | == | 4.6(2) | 0.0214(9) | 141.7 | 5932.88420 | -0.007 | $3\nu_2+\nu_3+\nu_4$ ($F_2$) | 3.30 | 0.05 |
| | R(3,$F_1$) | 5935.245195(3) | 9014.840901(3) | 0.74(3) | 1.36(9) | 4.409(8) | 0.64(1) | 126.3(5) | 5935.38477 | -0.14 | $3\nu_2+\nu_3+\nu_4$ ($F_2$) | 58.52 | 0.95 |

# Addendum to: Sub-Doppler optical-optical double-resonance spectroscopy using a cavity-enhanced frequency comb probe

**Vinicius Silva de Oliveira[1], Isak Silander[1], Lucile Rutkowski[2], Grzegorz Soboń[3], Ove Axner[1], Kevin K. Lehmann[4], and Aleksandra Foltynowicz[1,*]**

[1] Department of Physics, Umeå University, 901 87 Umeå, Sweden

[2] Univ Rennes, CNRS, IPR (Institut de Physique de Rennes)-UMR 6251, F-35000 Rennes, France

[3] Faculty of Electronics, Photonics and Microsystems, Wrocław University of Science and Technology, Wybrzeże Wyspiańskiego 27, 50-370 Wrocław, Poland

[4] Departments of Chemistry & Physics, University of Virginia, Charlottesville, VA 22904, USA

Corresponding author: aleksandra.foltynowicz@umu.se

In our Article[1], we assigned the experimental line list of hot-band methane transitions using the TheoReTS/HITEMP database[2], which at that time contained the most accurate predictions in the 9000 $cm^{-1}$ energy range. Soon after the publication of our Article, two other accurate theoretical methane line lists became available, namely the new ExoMol line list[3] and a line list obtained from an *ab initio*-based effective Hamiltonian[4]. In this Addendum, we compare our experimental line list to these new predictions. We believe this updated comparison is of wide interest to scientists working in e.g. astrophysics and combustion.

Moreover, we revisited the 45-times averaged R(2, $F_2$)–pumped spectrum, and we report four new weak lines from this spectrum, which are missing in TheoReTS but can be assigned using the ExoMol and the effective Hamiltonian line lists. We also removed the line at 5929.14547(1) $cm^{-1}$ from the list, because it was a false detection. This peak is a part of the instrumental line shape (ILS)[5] of the strong line at 5929.170466(3) $cm^{-1}$; it appeared in the 45-times averaged spectrum but was not visible in the spectrum averaged 5 times that was used for minimization of the ILS.

Most of the experimental lines could be unambiguously assigned to the ExoMol and Hamiltonian line lists based on the closest match to a strongest line. In some cases, patterns in intensity were also considered by matching lines whose relative ratios of experimental and predicted absorption coefficients were close to the mean value calculated from the unambiguously assigned lines, as described in the Article.

Addendum Table A1 is an updated version of the Supplementary Table 2, with the changes highlighted in bold. In this table, four new lines are reported, and the line at 5929.14547(1) $cm^{-1}$ has been removed. The vibrational assignments from the effective Hamiltonian are updated to match the newest calculations provided by Michael Rey based on Ref. [4].

Addendum Tables A2 and A3 contain the same experimental line list assigned using the ExoMol and Hamiltonian predictions, respectively.

Addendum Figure A1 is an updated version of Figure 6 from the Article. Panel a), which shows the integrated absorption coefficients of all detected lines, has been updated by the addition of 4 weak lines (marked by gray stars), and removal of one line. Panels b) and c), which show the ratios of the experimental and predicted integrated absorption coefficients, and the difference between the experimental and predicted line positions, remain unchanged. Addendum Figures A2 and A3 show the corresponding figures for the ExoMol and Hamiltonian assignments.

Addendum Table A4 summarizes the comparison of the experimental line lists to the three sets of predictions. It lists the mean values and standard deviations of the relative intensity ratios for the P(2, $F_2$)–, Q(2, $F_2$)–, and R(2, $F_2$)–pumped spectra shown in panels b) of Figures A1 – A3. The intensity ratios that are outliers in the R(2, $F_2$)–pumped spectra, marked by ♦ in Addendum Tables A1 – A3, are excluded from the calculations. Addendum Table A4 also lists the mean values and standard deviations of the differences between the experimental and the predicted transition wavenumbers, shown in panels c) of Figures A1 – A3. The new Hamiltonian predictions show the best agreement with the experimental data. This set of predictions was recently found to also yield the best agreement with experimental hot-band transitions reaching states with higher rotational quantum numbers than in this work[6].

**Acknowledgments**: The authors thank Michael Rey for providing the results of his recent calculations from the non-empirical effective Hamiltonian described in his work, Ref. [4], and Hiroyuki Sasada for providing the ground state term value from his unpublished work. A.F. acknowledges the Knut and Alice Wallenberg Foundation (KAW 2015.0159, KAW 2020.0303) and the Swedish Research Council (2020-00238). L.R. acknowledges the French National Research Agency (ANR-19-CE30-0038-01); G.S. acknowledges the Foundation for Polish Science (POIR.04.04.00-00-434D/17-00); O.A. acknowledges the Kempe Foundation (JCK 1317.1), K.K.L. acknowledges the National Science Foundation (grant: CHE-2108458).

**Addendum Table A1. Updated version of the Supplementary Table 2. Results of fits to all lines and comparison with the TheoReTS/HITEMP database.[2]** Column 1: Pump transition. Column 2: Assignment of the probe transition. Column 3: Experimental probe transition wavenumber. Column 4: Final state term value calculated as a sum of the term value for the initial ground vibrational state 2($F_2$) rotational state[7], the pump transition wavenumber[8,9] and the experimental probe transition wavenumber. Column 5: Integrated absorption coefficient of the Lorentzian part of the probe transition line shape. Column 6: Ratio of line intensities measured with parallel and perpendicular relative pump/probe polarizations. Column 7: Half width at half maximum (HWHM) of the Lorentzian part of the probe transition. Column 8: Integrated absorption coefficient of the Gaussian part. Column 9: HWHM of the Gaussian part. Column 10: Predicted probe transition wavenumber from TheoReTS/HITEMP. Column 11: Difference with respect to observed wavenumber (plotted in Figure 6c of the main paper and Addendum Figure A1c). Column 12: Final state dominant assignment from the Hamiltonian described in Ref. [4]. Column 13: Line intensity from TheoReTS/HITEMP at 296 K. Column 14: Integrated absorption coefficient from TheoReTS/HITEMP calculated at 50 mTorr and 296 K. The 9 transitions marked by an asterisk were detected only in the spectrum averaged 45 times with parallel pump/probe polarization. The changes in relation to Supplementary Table 2 are highlighted in bold. The transitions marked by ♦ are excluded from the relative intensity ratio calculations in Addendum Table A4.

| 1 | 2 | 3 | 4 | 5 | 6 | 7 | 8 | 9 | 10 | 11 | 12 | 13 | 14 |
|---|---|---|---|---|---|---|---|---|---|---|---|---|---|
| Pump transition in the $\nu_3$ band | Probe transition | Probe transition wavenumber [cm$^{-1}$] | Final state term value [cm$^{-1}$] | Probe transition integrated absorption coefficient [$10^{-9}$ cm$^{-2}$] | Polar. depend. intensity ratio | Probe transition width [MHz] | Gaussian integrated absorption [$10^{-9}$ cm$^{-2}$] | Gaussian width [MHz] | TheoReTS transition wavenumber [cm$^{-1}$] | Obs. – pred. wavenum. [cm$^{-1}$] | Final state assignment from Ref. [4]. | TheoReTS transition intensity [$10^{-30}$ cm/molec] | TheoReTS integrated absorption coefficient [$10^{-13}$ cm$^{-2}$] |
| P(2,$F_2$) | R(1) | 5948.267590(3) | 8978.704010(3) | 2.26(9) | 0.91(5) | 5.093(8) | 1.50(3) | 144.3(4) | 5947.67695 | 0.59 | $3\nu_3$ ($F_1$) | 220.40 | 3.59 |
| | R(1) | 5964.06227(2) | 8994.49869(2) | 0.081(4) | 1.01(8) | 6.5(3) | 0.058(2) | 141.7 | 5965.37197 | -1.31 | **$5\nu_2+\nu_4$ ($F_1$)** | 7.79 | 0.13 |
| | R(1) | 5979.042972(3) | 9009.479392(3) | 0.56(2) | 0.98(6) | 4.459(8) | 0.396(9) | 125.5(6) | 5979.71721 | -0.67 | $\nu_1+4\nu_2$ ($A_1$) | 68.70 | 1.12 |
| Q(2,$F_2$) | Q(2,$F_1$) | 5928.61142(2) | 8978.70401(2) | 0.56(4) | 1.5(3) | 6.0(1) | 0.40(2) | 141.7 | 5928.02556 | 0.59 | $3\nu_3$ ($F_1$) | 53.28 | 0.87 |
| | Q(2,$F_1$) | 5944.40608(2) | 8994.49868(2) | 0.103(5) | 1.5(3) | 4.31(9) | 0.074(2) | 141.7 | 5945.72057 | -1.31 | **$5\nu_2+\nu_4$ ($F_1$)** | 10.83 | 0.18 |
| | R(2,$F_1$) | 5958.673574(6) | 9008.766169(6) | 2.38(10) | 0.83(7) | 4.73(8) | 2.20(8) | 143(4) | 5957.83620 | 0.84 | $3\nu_3$ ($F_1$) | 243.60 | 3.97 |
| | Q(2,$F_1$) | 5959.386797(5) | 9009.479392(5) | 0.89(3) | 1.8(3) | 4.29(1) | 0.64(1) | 141.7 | 5960.06582 | -0.68 | $\nu_1+4\nu_2$ ($A_1$) | 82.82 | 1.35 |
| R(2,$F_2$) | ***==** | **5909.91517(2)** | **8989.51088(2)** | **0.016(2)** | == | **4.8(6)** | **0.011(2)** | **141.7** | == | == | == | == | == |
| | R(3, $F_1$) | 5913.18732(2) | 8992.78303(2) | 0.062(4) | 1.1(1) | 3.6(3) | 0.045(3) | 141.7 | 5913.97472 | -0.79 | **$3\nu_2+\nu_3+\nu_4$ ($F_1$)** | 6.48 | 0.11 |
| | *== | **5913.90237(3)** | **8993.49807(3)** | **0.007(1)** | == | **3.6(6)** | **0.0050(7)** | **141.7** | == | == | == | == | == |
| | R(3,$F_1$) | 5918.14141(1) | 8997.73712(1) | 0.096(5) | 1.05(10) | 4.1(1) | 0.069(3) | 141.7 | 5918.79890 | -0.66 | $3\nu_2+\nu_3+\nu_4$ ($F_2$) | 11.14 | 0.18 |
| | *R(3,$F_1$) | 5920.775507(9) | 9000.371213(9) | 0.039(2) | == | 5.67(6) | 0.034(1) | 141.7 | 5921.56979 | -0.79 | **$3\nu_2+\nu_3+\nu_4$ (E)** | 4.43 | 0.07 |
| | *== | 5922.51469(4) | 9002.11039(4) | 0.012(2) | == | 4(1) | 0.010(2) | 141.7 | == | == | == | == | == |
| | *♦R(3,$F_1$) | 5923.32503(5) | 9002.92074(4) | 0.039(6) | == | 9(2) | 0.033(4) | 141.7 | 5923.25121 | 0.074 | **$3\nu_2+\nu_3+\nu_4$ ($F_1$)** | 2.17 | 0.04 |
| | R(3,$F_1$) | 5923.94848(1) | 9003.54418(1) | 0.44(2) | 1.19(10) | 5.2(2) | 0.32(1) | 141.7 | 5924.39254 | -0.44 | **$3\nu_2+\nu_3+\nu_4$ ($F_2$)** | 49.03 | 0.80 |
| | R(3,$F_1$) | 5924.26536(2) | 9003.86107(2) | 0.175(7) | 1.1(1) | 3.79(8) | 0.126(3) | 141.7 | 5924.10232 | 0.16 | **$5\nu_2+\nu_4$ ($F_1$)** | 13.00 | 0.21 |
| | *== | **5924.73759(3)** | **9004.33329(3)** | **0.012(3)** | == | **4.2(10)** | **0.008(2)** | **141.7** | == | == | == | == | == |
| | *♦R(3,$F_1$) | 5924.99660(1) | 9004.59230(1) | 0.049(4) | == | 4.3(3) | 0.041(2) | 141.7 | 5924.99972 | -0.003 | $3\nu_2+\nu_3+\nu_4$ ($F_2$) | 2.23 | 0.04 |
| | Q(3,$F_1$) | 5929.170466(3) | 9008.766172(3) | 4.7(2) | 0.4(1) | 5.87(1) | 2.81(7) | 142.4(3) | 5928.34007 | 0.83 | $3\nu_3$ ($F_1$) | 418.90 | 6.83 |
| | P(3,$F_1$) | 5929.883687(4) | 9009.479393(4) | 0.85(3) | 1.6(2) | 4.51(2) | 0.61(1) | 141.7 | 5930.56969 | -0.69 | $\nu_1+4\nu_2$ ($A_1$) | 71.45 | 1.16 |
| | R(3,$F_1$) | 5932.279186(9) | 9011.874892(9) | 0.100(4) | 1.2(1) | 4.00(6) | 0.072(2) | 141.7 | 5932.52779 | -0.25 | **$\nu_1+2\nu_2+\nu_3$ ($F_2$)** | 8.00 | 0.13 |
| | *R(3,$F_1$) | 5932.876751(8) | 9012.472457(8) | 0.025(2) | == | 4.6(2) | 0.0214(9) | 141.7 | 5932.88420 | -0.007 | $3\nu_2+\nu_3+\nu_4$ ($F_2$) | 3.30 | 0.05 |
| | R(3,$F_1$) | 5935.245195(3) | 9014.840901(3) | 0.74(3) | 1.36(9) | 4.409(8) | 0.64(1) | 126.3(5) | 5935.38477 | -0.14 | $3\nu_2+\nu_3+\nu_4$ ($F_2$) | 58.52 | 0.95 |
| | *== | **5936.15774(1)** | **9015.75344(1)** | **0.019(1)** | == | **7.5(2)** | **0.0135(9)** | **141.7** | == | == | == | == | == |

**Addendum Table A2. Comparison with the ExoMol database.**[3] Column 1: Pump transition. Column 2: Assignment of the probe transition. Column 3: Experimental probe transition wavenumber. Column 4: Experimental final state term value. Column 5: Integrated absorption coefficient of the probe transition. Column 6: Predicted probe transition wavenumber from ExoMol. Column 7: Difference with respect to observed wavenumber (Addendum Figure A2c). Column 8: Final state dominant assignment from ExoMol. Column 9: Line intensity from ExoMol at 296 K. Column 10: Integrated absorption from ExoMol calculated at 50 mTorr and 296 K. Column 11: Polyad counting number. The 9 transitions marked by an asterisk were detected only in the spectrum averaged 45 times with parallel pump/probe polarization. The transitions marked by ♦ are excluded from the relative intensity ratio calculations shown in Addendum Table A4.

| 1 | 2 | 3 | 4 | 5 | 6 | 7 | 8 | 9 | 10 | 11 |
|---|---|---|---|---|---|---|---|---|---|---|
| Pump transition in the $\nu_3$ band | Probe transition | Probe transition wavenumber [cm$^{-1}$] | Final state term value [cm$^{-1}$] | Probe transition integrated absorption coefficient [10$^{-9}$ cm$^{-2}$] | ExoMol transition wavenumber [cm$^{-1}$] | Obs. –pred. wavenum. [cm$^{-1}$] | Final state assignment from ExoMol. | ExoMol transition intensity [10$^{-30}$ cm/molec] | ExoMol integrated absorption coefficient [10$^{-13}$ cm$^{-2}$] | Polyad counting number |
| P(2,F$_2$) | R(1) | 5948.267590(3) | 8978.704010(3) | 2.26(9) | 5949.35158 | -1.08 | $3\nu_3$ (F$_1$) | 271.49 | 4.43 | 386 |
| | R(1) | 5964.06227(2) | 8994.49869(2) | 0.081(4) | 5964.85821 | -0.80 | $5\nu_2+\nu_4$ (F$_1$) | 6.86 | 0.11 | 391 |
| | R(1) | 5979.042972(3) | 9009.479392(3) | 0.56(2) | 5979.51580 | -0.47 | $\nu_1+2\nu_3$ (A$_1$) | 67.98 | 1.11 | 393 |
| Q(2,F$_2$) | Q(2,F$_1$) | 5928.61142(2) | 8978.70401(2) | 0.56(4) | 5929.69541 | -1.08 | $3\nu_3$ (F$_1$) | 50.90 | 0.83 | 386 |
| | Q(2,F$_1$) | 5944.40608(2) | 8994.49868(2) | 0.103(5) | 5945.20204 | -0.80 | $5\nu_2+\nu_4$ (F$_1$) | 9.57 | 0.16 | 391 |
| | R(2,F$_1$) | 5958.673574(6) | 9008.766169(6) | 2.38(10) | 5959.76805 | -1.09 | $3\nu_3$ (F$_1$) | 246.39 | 4.02 | 520 |
| | Q(2,F$_1$) | 5959.386797(5) | 9009.479392(5) | 0.89(3) | 5959.85963 | -0.47 | $\nu_1+2\nu_3$ (A$_1$) | 81.18 | 1.32 | 393 |
| R(2,F$_2$) | *R(3,F$_1$) | 5909.91517(2) | 8989.51088(2) | 0.016(2) | 5910.63376 | -0.72 | $3\nu_2+\nu_3+\nu_4$ (E) | 1.56 | 0.03 | 654 |
| | R(3,F$_1$) | 5913.18732(2) | 8992.78303(2) | 0.062(4) | 5913.81192 | -0.62 | $5\nu_2+\nu_4$ (F$_2$) | 6.43 | 0.10 | 656 |
| | *R(3,F$_1$) | 5913.90237(3) | 8993.49807(3) | 0.007(1) | 5914.36797 | -0.47 | $5\nu_2+\nu_4$ (F$_2$) | 0.74 | 0.01 | 657 |
| | R(3,F$_1$) | 5918.14141(1) | 8997.73712(1) | 0.096(5) | 5918.64464 | -0.50 | $5\nu_2+\nu_4$ (F$_2$) | 7.71 | 0.13 | 661 |
| | *R(3,F$_1$) | 5920.775507(9) | 9000.371213(9) | 0.039(2) | 5921.17779 | -0.40 | $\nu_1+2\nu_2+\nu_3$ (F$_2$) | 4.24 | 0.07 | 663 |
| | *♦R(3,F$_1$) | 5922.51469(4) | 9002.11039(4) | 0.012(2) | 5922.51575 | -0.001 | $3\nu_2+\nu_3+\nu_4$ (F$_2$) | 0.72 | 0.01 | 664 |
| | *♦R(3,F$_1$) | 5923.32503(5) | 9002.92074(4) | 0.039(6) | 5923.84294 | -0.52 | $3\nu_2+\nu_3+\nu_4$ (F$_1$) | 2.14 | 0.03 | 665 |
| | R(3,F$_1$) | 5923.94848(1) | 9003.54418(1) | 0.44(2) | 5924.36897 | -0.42 | $5\nu_2+\nu_4$ (F$_2$) | 43.29 | 0.71 | 666 |
| | R(3,$F_1$) | 5924.26536(2) | 9003.86107(2) | 0.175(7) | 5924.82589 | -0.56 | 5$\nu_2$ +$\nu_4$ ($F_2$) | 17.48 | 0.28 | 667 |
| | *♦R(3,$F_1$) | 5924.73759(3) | 9004.33329(3) | 0.012(3) | 5925.15880 | -0.42 | 3$\nu_3$ ($F_2$) | 0.36 | 0.01 | 668 |
| | *♦R(3,$F_1$) | 5924.99660(1) | 9004.59230(1) | 0.049(4) | 5925.52906 | -0.53 | 3$\nu_2$+$\nu_3$+$\nu_4$($F_1$) | 1.23 | 0.02 | 669 |
| | Q(3,$F_1$) | 5929.170466(3) | 9008.766172(3) | 4.7(2) | 5930.26494 | -1.09 | 3$\nu_3$($F_1$) | 422.39 | 6.89 | 520 |
| | P(3,$F_1$) | 5929.883687(4) | 9009.479393(4) | 0.85(3) | 5930.35652 | -0.47 | $\nu_1$+2$\nu_3$($A_1$) | 68.81 | 1.12 | 393 |
| | R(3,$F_1$) | 5932.279186(9) | 9011.874892(9) | 0.100(4) | 5932.28078 | -0.002 | 3$\nu_3$ ($F_2$) | 10.32 | 0.17 | 675 |
| | *R(3,$F_1$) | 5932.876751(8) | 9012.472457(8) | 0.025(2) | 5932.88395 | -0.01 | 3$\nu_3$ ($F_2$) | 2.94 | 0.05 | 676 |
| | R(3,$F_1$) | 5935.245195(3) | 9014.840901(3) | 0.74(3) | 5935.35204 | -0.11 | 3$\nu_3$ ($F_2$) | 71.99 | 1.17 | 677 |
| | *R(3,$F_1$) | 5936.15774(1) | 9015.75344(1) | 0.019(1) | 5936.26181 | -0.10 | 3$\nu_3$ ($F_2$) | 1.27 | 0.02 | 678 |

**Addendum table A3. Comparison with the effective Hamiltonian model.**[4] Column 1: Pump transition. Column 2: Assignment of the probe transition. Column 3: Experimental probe transition wavenumber. Column 4: Experimental final state term value. Column 5: Integrated absorption coefficient of the probe transition. Column 6: Predicted probe transition wavenumber from the Hamiltonian. Column 7: Difference with respect to observed wavenumber (Addendum Figure A3c). Column 8: Final state dominant assignment from the Hamiltonian. Column 9: Line intensity from the Hamiltonian at 296 K. Column 10: Integrated absorption coefficient from Hamiltonian calculated at 50 mTorr and 296 K. Column 11: Polyad counting number. The 9 transitions marked by an asterisk were detected only in the spectrum averaged 45 times with parallel pump/probe polarization. The transitions marked by ♦ are excluded from the relative intensity ratio calculations shown in Addendum Table A4.

| 1 | 2 | 3 | 4 | 5 | 6 | 7 | 8 | 9 | 10 | 11 |
|---|---|---|---|---|---|---|---|---|---|---|
| Pump transition in the $\nu_3$ band | Probe transition | Probe transition wavenumber [$cm^{-1}$] | Final state term value [$cm^{-1}$] | Probe transition integrated absorption coefficient [$10^{-9}$ $cm^{-2}$] | Hamiltonian transition wavenumber [$cm^{-1}$] | Obs. – pred. wavenum. [$cm^{-1}$] | Final state assignment from Hamiltonian | Hamiltonian transition intensity [$10^{-30}$ cm/molec] | Hamiltonian integrated absorption coefficient [$10^{-13}$ $cm^{-2}$] | Polyad counting number |
| P(2,$F_2$) | R(1) | 5948.267590(3) | 8978.704010(3) | 2.26(9) | 5948.32948 | -0.06 | $3\nu_3$ ($F_1$) | 271.9 | 4.43 | 386 |
| | R(1) | 5964.06227(2) | 8994.49869(2) | 0.081(4) | 5964.25651 | -0.19 | $5\nu_2+\nu_4$ ($F_1$) | 7.83 | 0.13 | 391 |
| | R(1) | 5979.042972(3) | 9009.479392(3) | 0.56(2) | 5978.95570 | 0.09 | $\nu_1+4\nu_2$ ($A_1$) | 69.39 | 1.13 | 393 |
| Q(2,$F_2$) | Q(2,$F_1$) | 5928.61142(2) | 8978.70401(2) | 0.56(4) | 5928.67350 | -0.06 | $3\nu_3$ ($F_1$) | 54.91 | 0.90 | 386 |
| | Q(2,$F_1$) | 5944.40608(2) | 8994.49868(2) | 0.103(5) | 5944.60053 | -0.19 | $5\nu_2+\nu_4$ ($F_1$) | 10.45 | 0.17 | 391 |
| | R(2,$F_1$) | 5958.673574(6) | 9008.766169(6) | 2.38(10) | 5958.71365 | -0.04 | $3\nu_3$ ($F_1$) | 245.2 | 4.00 | 520 |
| | Q(2,$F_1$) | 5959.386797(5) | 9009.479392(5) | 0.89(3) | 5959.29972 | 0.09 | $\nu_1+4\nu_2$ ($A_1$) | 83.73 | 1.37 | 393 |
| R(2,$F_2$) | *R(3,$F_1$) | 5909.91517(2) | 8989.51088(2) | 0.016(2) | 5910.23702 | -0.32 | $3\nu_2+\nu_3+\nu_4$ (E) | 2.09 | 0.03 | 654 |
| | R(3,$F_1$) | 5913.18732(2) | 8992.78303(2) | 0.062(4) | 5913.51328 | -0.33 | $3\nu_2+\nu_3+\nu_4$ ($F_1$) | 6.30 | 0.10 | 656 |
| | *R(3,$F_1$) | 5913.90237(3) | 8993.49807(3) | 0.007(1) | 5914.17819 | -0.28 | $3\nu_2+\nu_3+\nu_4$ ($F_1$) | 0.97 | 0.02 | 657 |
| | R(3,$F_1$) | 5918.14141(1) | 8997.73712(1) | 0.096(5) | 5918.31511 | -0.17 | $3\nu_2+\nu_3+\nu_4$ ($F_2$) | 9.54 | 0.16 | 661 |
| | *R(3,$F_1$) | 5920.775507(9) | 9000.371213(9) | 0.039(2) | 5920.95047 | -0.17 | $3\nu_2+\nu_3+\nu_4$ (E) | 5.65 | 0.09 | 663 |
| | *R(3,$F_1$) | 5922.51469(4) | 9002.11039(4) | 0.012(2) | 5922.51552 | -0.001 | $3\nu_2+\nu_3+\nu_4$ ($F_2$) | 0.82 | 0.01 | 664 |
| | *R(3,$F_1$) | 5923.32503(5) | 9002.92074(4) | 0.039(6) | 5923.42202 | -0.10 | $3\nu_2+\nu_3+\nu_4$ ($F_1$) | 3.13 | 0.05 | 665 |
| | R(3,$F_1$) | 5923.94848(1) | 9003.54418(1) | 0.44(2) | 5923.90046 | 0.05 | $3\nu_2+\nu_3+\nu_4$ ($F_2$) | 44.78 | 0.73 | 666 |
| | ♦R(3,$F_1$) | 5924.26536(2) | 9003.86107(2) | 0.175(7) | 5924.03473 | 0.23 | $5\nu_2+\nu_4$ ($F_1$) | 1.20 | 0.02 | 667 |
| | *R(3,$F_1$) | 5924.73759(3) | 9004.33329(3) | 0.012(3) | 5924.72996 | 0.01 | $3\nu_2+\nu_3+\nu_4$ ($F_2$) | 2.47 | 0.04 | 668 |
| | *R(3,$F_1$) | 5924.99660(1) | 9004.59230(1) | 0.049(4) | 5925.05501 | -0.06 | $3\nu_2+\nu_3+\nu_4$ ($F_2$) | 8.02 | 0.13 | 669 |
| | Q(3,$F_1$) | 5929.170466(3) | 9008.766172(3) | 4.7(2) | 5929.20981 | -0.04 | $3\nu_3$ ($F_1$) | 422.50 | 6.89 | 520 |
| | P(3,$F_1$) | 5929.883687(4) | 9009.479393(4) | 0.85(3) | 5929.79588 | 0.09 | $\nu_1+4\nu_2$ ($A_1$) | 71.65 | 1.17 | 393 |
| | R(3,$F_1$) | 5932.279186(9) | 9011.874892(9) | 0.100(4) | 5932.28640 | -0.01 | $\nu_1+2\nu_2+\nu_3$ ($F_2$) | 9.23 | 0.15 | 675 |
| | *R(3,$F_1$) | 5932.876751(8) | 9012.472457(8) | 0.025(2) | 5932.88420 | -0.01 | $3\nu_2+\nu_3+\nu_4$ ($F_2$) | 2.65 | 0.04 | 676 |
| | R(3,$F_1$) | 5935.245195(3) | 9014.840901(3) | 0.74(3) | 5935.22281 | 0.02 | $3\nu_2+\nu_3+\nu_4$ ($F_2$) | 63.29 | 1.03 | 677 |
| | *♦R(3,$F_1$) | 5936.15774(1) | 9015.75344(1) | 0.019(1) | 5936.23057 | -0.07 | $3\nu_2+\nu_3+\nu_4$ ($F_2$) | 1.03 | 0.02 | 678 |

**Addendum Table A4. Comparison of the observed and predicted transitions intensities and wavenumbers.** Column 1: Reference line list. Columns 2 to 7: mean values and standard deviations of the relative intensity ratios excluding outliers for P(2, $F_2$)-, Q(2, $F_2$)- and R(2, $F_2$)-pumped spectra, respectively. Columns 8 and 9: Mean values and standard deviations of differences between observed and predicted wavenumbers of all assigned lines.

| Reference | Relative intensity ratios | | | | | | Center wavenumbers [cm$^{-1}$] | |
|---|---|---|---|---|---|---|---|---|
| | P(2, $F_2$) | | Q(2, $F_2$) | | R(2, $F_2$) | | | |
| | Mean ratio | Standard deviation | Mean ratio | Standard deviation | Mean ratio | Standard deviation | Mean offset | Standard deviation |
| TheoReTS/HITEMP[2] | 5900 | 700 | 6200 | 400 | 6400 | 1200 | -0.24 | 0.65 |
| ExoMol[3] | 5800 | 1200 | 6500 | 390 | 6500 | 1100 | -0.58 | 0.34 |
| Hamiltonian[4] | 5500 | 740 | 6200 | 250 | 5900 | 1700 | -0.04 | 0.13 |

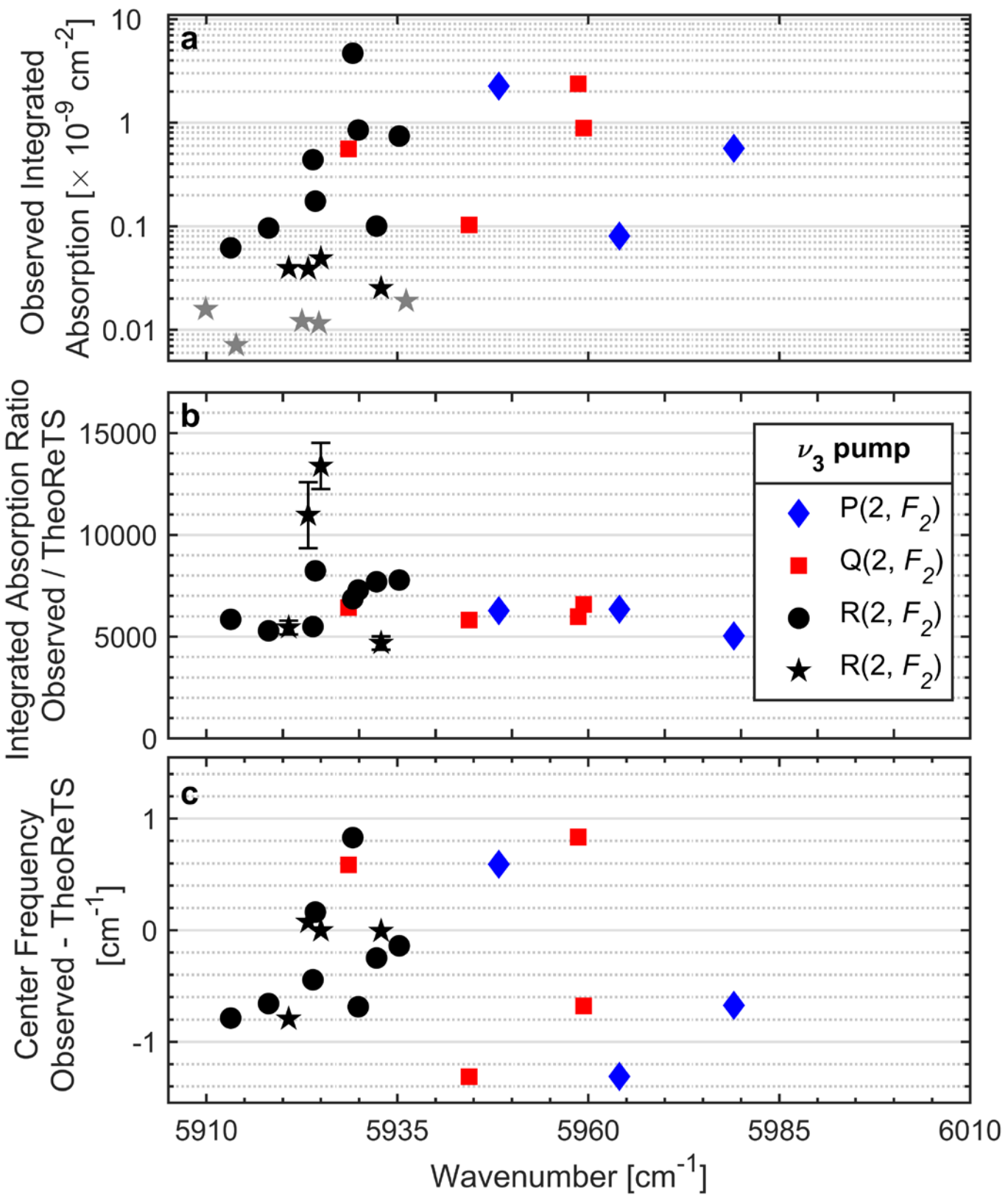

**Addendum Figure A1 a** Integrated absorption coefficient of all measured OODR probe transitions on a logarithmic scale. The markers show different pump transitions in the n3 band, and the stars show transitions visible only in the spectrum averaged 45 times with pump locked to the n3 R(2, $F_2$) transition. The gray stars indicate transitions that lack assignment in TheoReTS/HITEMP. **b** Ratios of the experimental intergrated absorption coefficients from **a** to integrated absorption coefficients predicted at 296 K and 50 mTorr from the TheoReTS/HITEMP database. **c** Center wavenumbers of the OODR probe transitions compared to predictions from the TheoReTS/HITEMP database.

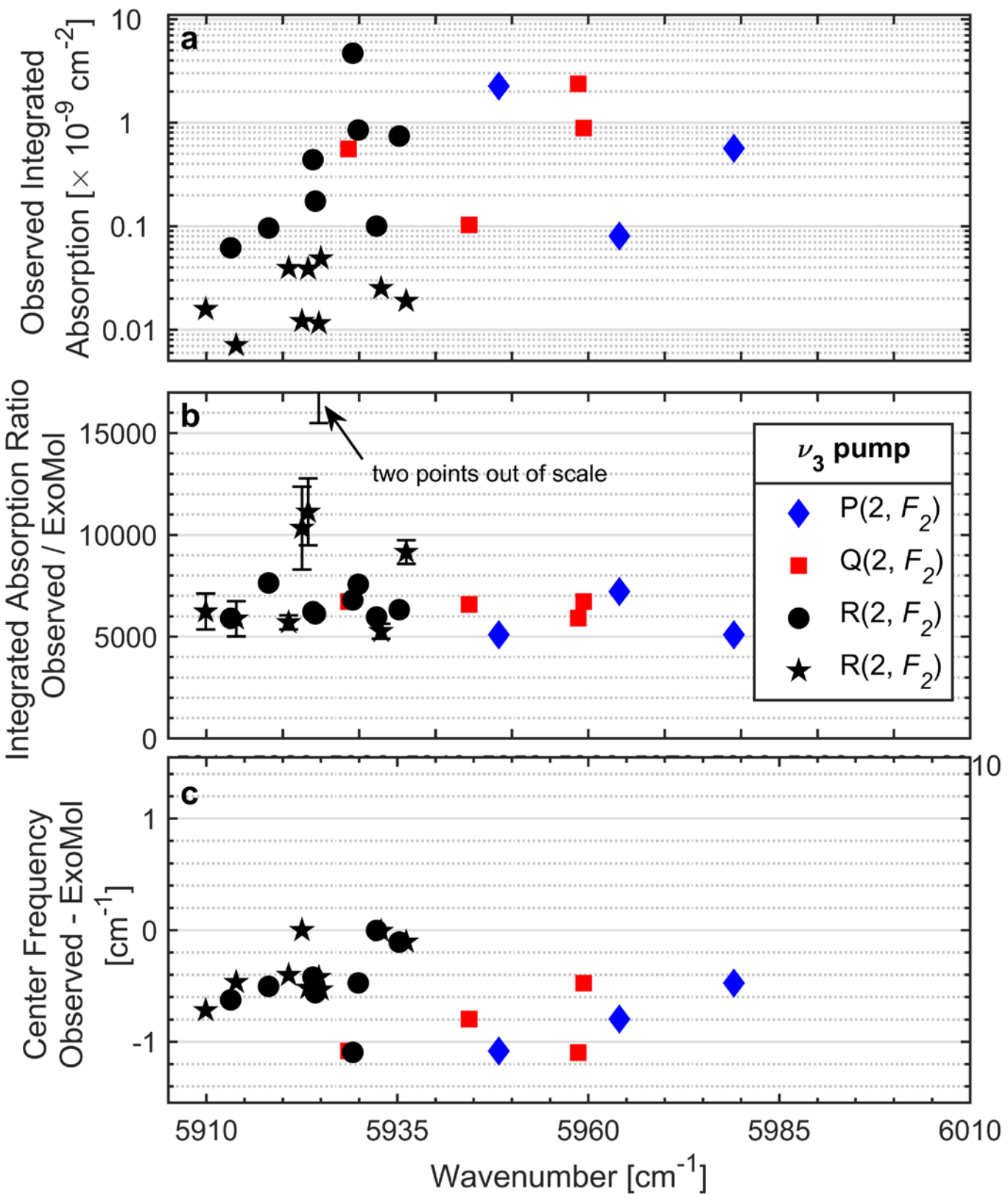

**Addendum Figure A2** **a** Integrated absorption coefficient of all measured OODR probe transitions on a logarithmic scale. **b** Ratios of the experimental intergrated absorption coefficients from **a** to integrated absorption coefficients predicted at 296 K and 50 mTorr from the ExoMol database. **c** Center wavenumbers of the OODR probe transitions compared to predictions from the ExoMol database. The markers show different pump transitions in the $n_3$ band, and the stars show the transitions detected in the spectrum averaged 45 times with pump locked to $n_3$ R(2, $F_2$) transition.

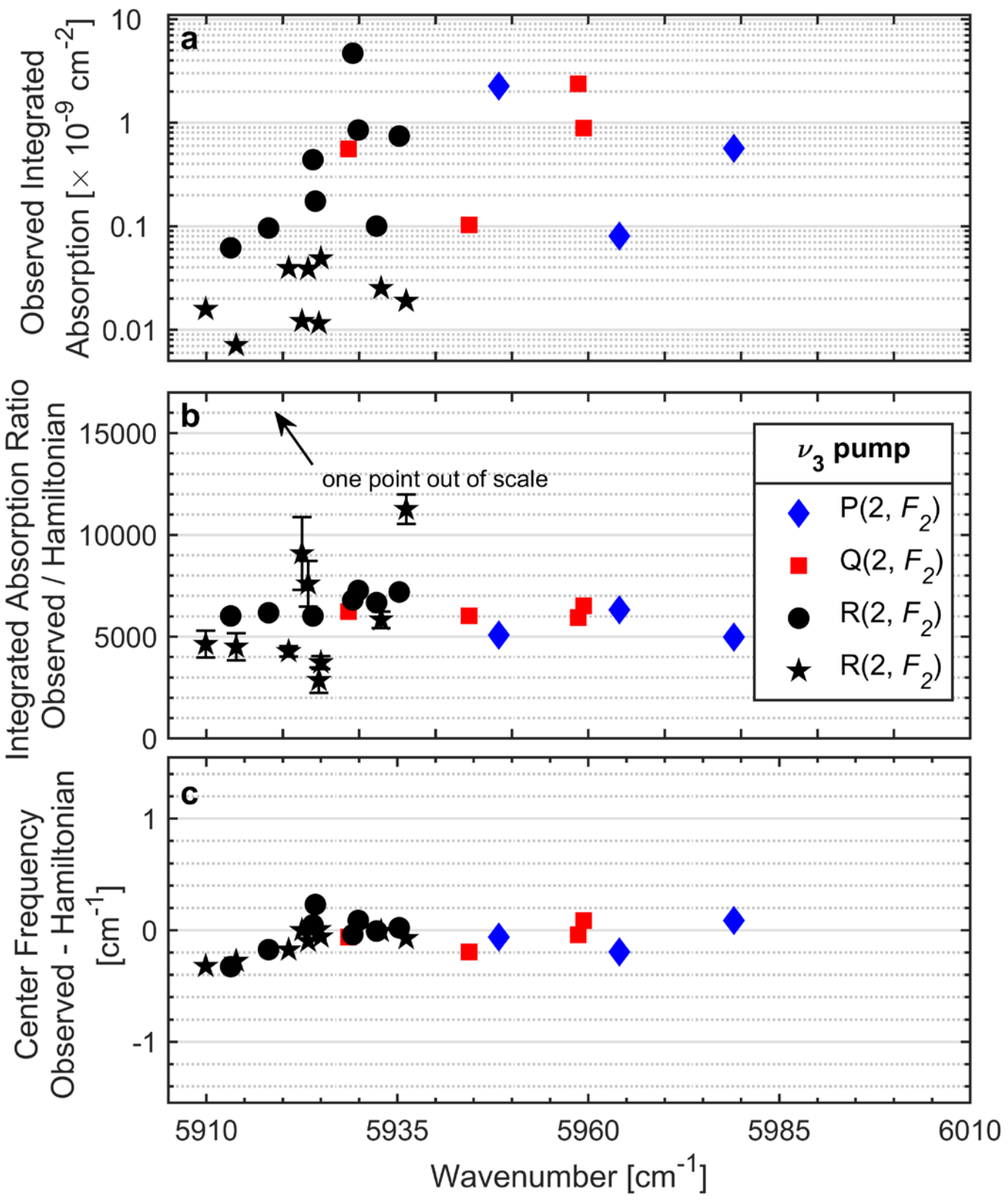

**Addendum Figure A3 a** Integrated absorption coefficient of all measured OODR probe transitions on a logarithmic scale. **b** Ratios of the experimental intergrated absorption coefficients from **a** to integrated absorption coefficients predicted at 296 K and 50 mTorr from the Hamiltonian line list. **c** Center wavenumbers of the OODR probe transitions compared to predictions from the Hamiltonian database. The markers show different pump transitions in the $n_3$ band, and the stars show the transitions detected in the spectrum averaged 45 times with pump locked to $n_3$ R(2, $F_2$) transition.